\documentclass[aps,prl,twocolumn,superscriptaddress,amsmath,nofootinbib,10pt,floatfix]{revtex4-2}

\usepackage[english]{babel}
\usepackage{graphicx}
\usepackage{ulem}
\usepackage{soul}
\usepackage{float} 
\usepackage[colorlinks, linkcolor=red , citecolor= blue,breaklinks=true]{hyperref}
\usepackage{bm}
\usepackage{verbatim}
\usepackage{esint}
\usepackage{marginnote}
\usepackage{feynmf}

\renewcommand{\vec}[1]{\boldsymbol{#1}}
\newcommand{\RNum}[1]{\uppercase\expandafter{\romannumeral #1\relax}}

\def \ve {\varepsilon}

\def \v {{\bf v}}

\def \ve {\varepsilon}

\def \beq {\begin{eqnarray}}
\def \eeq {\end{eqnarray}}
\def \tn {\textnormal}

\def \tn {\textnormal}
\def \nn {\nonumber}

\begin{document}
\title{$T-$linear resistivity from magneto-elastic scattering: application to PdCrO$_2$}

\author{J.F. Mendez-Valderrama} \thanks{These authors contributed equally to this work}
\affiliation{Department of Physics, Cornell University, Ithaca, New York 14853, USA.}
\author{Evyatar Tulipman} \thanks{These authors contributed equally to this work}
\affiliation{Department of Condensed Matter Physics, Weizmann Institute of Science, Rehovot 76100, Israel}
\author{Elina Zhakina}
\affiliation{Max Planck Institute for Chemical Physics of Solids,
N\"othnitzer Stra\ss e 40, 01187 Dresden, Germany}
\author{Andrew P. Mackenzie}
\affiliation{Max Planck Institute for Chemical Physics of Solids,
N\"othnitzer Stra\ss e 40, 01187 Dresden, Germany}
\affiliation{Scottish Universities Physics Alliance, School of Physics \& Astronomy, University of St. Andrews, St. Andrews KY16 9SS, United Kingdom}
\author{Erez Berg}
\affiliation{Department of Condensed Matter Physics, Weizmann Institute of Science, Rehovot 76100, Israel}
\author{Debanjan Chowdhury}
\affiliation{Department of Physics, Cornell University, Ithaca, New York 14853, USA.}

\date{\today}

\begin{abstract}
An electronic solid with itinerant carriers and localized magnetic moments represents a paradigmatic strongly correlated system. The electrical transport properties associated with the itinerant carriers, as they scatter off these local moments, has been scrutinized across a number of materials. Here we analyze the transport characteristics associated with ultra-clean PdCrO$_2$ --- a quasi two-dimensional material consisting of alternating layers of itinerant Pd-electrons and Mott-insulating CrO$_2$ layers --- which shows a pronounced regime of $T-$linear resistivity over a wide-range of intermediate temperatures. By contrasting these observations to the transport properties in a closely related material PdCoO$_2$, where the CoO$_2$ layers are band-insulators, we can rule out the traditional electron-phonon interactions as being responsible for this interesting regime. We propose a previously ignored electron-magnetoelastic interaction between the Pd-electrons, the Cr local-moments and an out-of-plane phonon as the main scattering mechanism that leads to the significant enhancement of resistivity and a $T-$linear regime in PdCrO$_2$ at temperatures far in excess of the magnetic ordering temperature. We suggest a number of future experiments to confirm this picture in PdCrO$_2$, as well as other layered metallic/Mott-insulating materials.     
\end{abstract}

\maketitle

{\it Introduction.-} Recent years have witnessed a resurgence of interest in the microscopic origin of an electrical resistivity that scales linearly with temperature \cite{chowdhury_sachdev-ye-kitaev_2022,varma_colloquium_2020} and exhibiting a Planckian scattering rate, $\Gamma= Ck_BT/\hbar$, where $C\sim O(1)$ coefficient \cite{bruin_similarity_2013,hartnoll_colloquium_2022,TBG2019,legros_universal_2019,grissonnanche_linear-temperature_2021}. In conventional (simple) metals at room temperature, this phenomenology is readily understood as a consequence of electrons scattering off thermally excited phonons in their equipartition regime \cite{Ziman_2001}. On the other hand, numerous ``correlated" materials belonging to the cuprate  \cite{takagi_systematic_1992,hussey_dichotomy_2011,legros_universal_2019,grissonnanche_linear-temperature_2021}, pnictide \cite{Doiron2009,Shibauchi2014}, ruthenates \cite{Allen1996,klein_anomalous_1996,hussey_normal-state_1998,schneider_low-energy_2014,Rost2011}, rare-earth \cite{stewart_non-fermi-liquid_2001} and moir\'e bilayers \cite{TBG2019,jaoui_quantum_2022} display Planckian scattering down to low temperatures, likely driven by purely electronic interactions and where {\it a priori} it is unclear if phonons play an essential role \cite{bruin_similarity_2013,sds19,das_sarma_strange_2022,hartnoll_planckian_2022}. It is challenging to disentangle the role of electron-electron and electron-phonon interactions on scattering lifetimes. It is quite natural to ask if materials with a nearly identical phonon spectrum and distinct electronic spectra can lead to a distinct temperature dependence of their respective resistivities. 
\begin{figure}[t]
\centering
\includegraphics[width=0.5\textwidth]{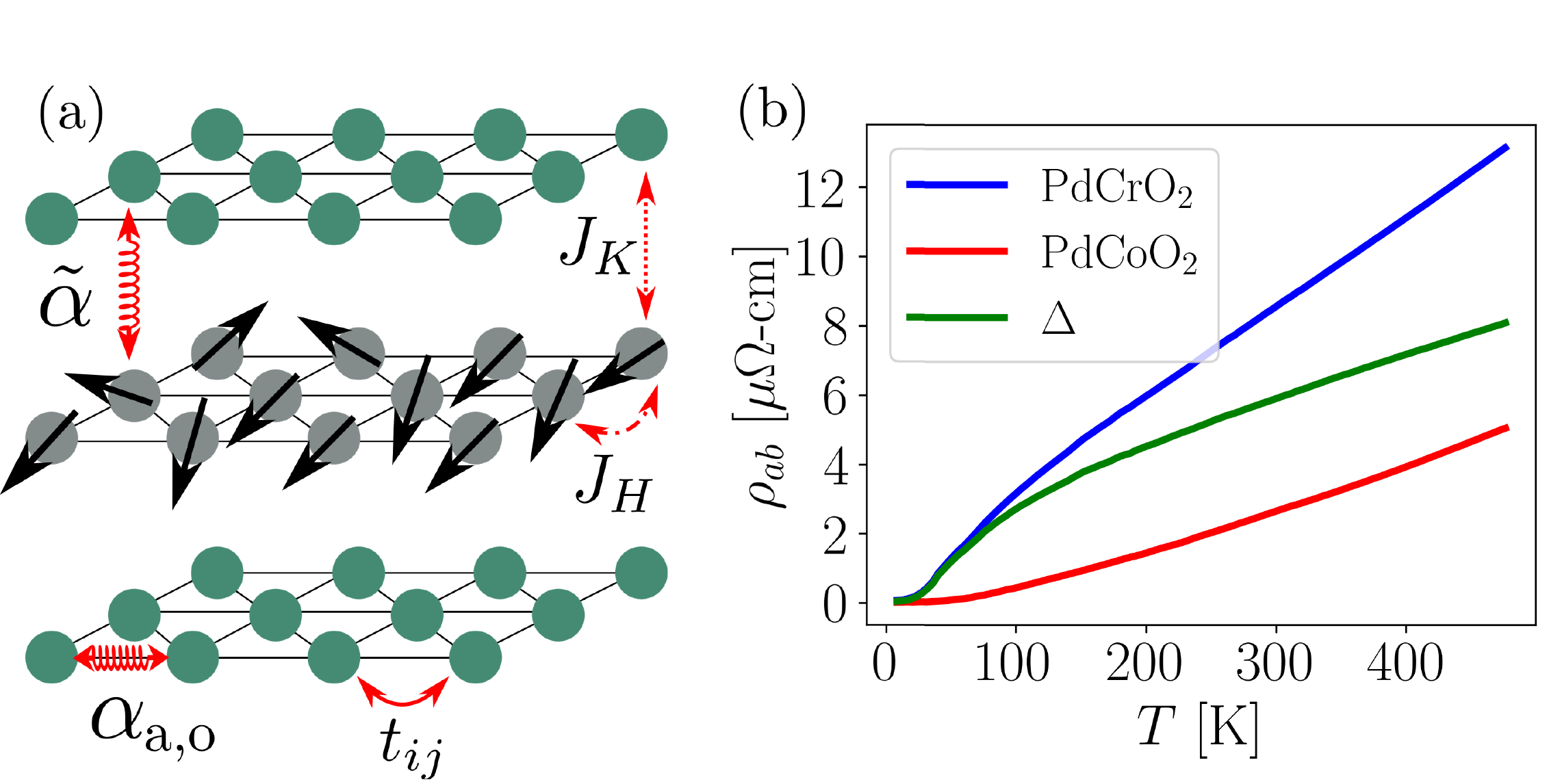} 
\caption{(a)  The structural motif in PdCrO$_2$, composed of alternating layers of triangular lattices of conducting Pd planes (green) and Mott insulating CrO$_2$ planes (grey). The different coupling constants denoted in the figure are as defined in Eq.~\eqref{total_H}. (b) The in-plane resistivities of PdCrO$_2$ and PdCoO$_2$ taken from Ref.~\cite{Hicks2015} along with their difference $\Delta \rho \equiv \rho_{ab}^{\rm PdCrO_2} -\rho_{ab}^{\rm PdCoO_2} >0$.}
\label{fig:model}
\end{figure}

\allowdisplaybreaks{
{\it An experimental puzzle.-} The goal of this letter is to resolve a conundrum inspired by electrical transport measurements in two isostructural quasi two-dimensional compounds with distinct electronic structures: PdCoO$_2$ and PdCrO$_2$. Their structural motif consists of alternately stacked layers of highly conducting Pd and insulating CoO$_2$/CrO$_2$ in a triangular lattice arrangement \cite{eyert_metallic_2008,takatsu_critical_2009,ong_origin_2010,mackenzie_properties_2017,sunko_probing_2020}; see Fig.~\ref{fig:model}(a). The phonon spectra for the two compounds are nearly identical; some of the differences arise from the distinct ionic masses \cite{takatsu_roles_2007,glamazda_collective_2014, mackenzie_properties_2017,sunko_probing_2020}, and recent analysis has also revealed that the unit cell of PdCrO$_2$ is slightly enlarged due to the presence of magnetic moments \cite{Zhakina2023}. On the other hand, their electronic spectra are different since the CrO$_2$ layers are Mott-insulating with the local-moments interacting via antiferromagnetic (AFM) exchange interactions \cite{takatsu_critical_2009,takatsu_magnetic_2014}, while the CoO$_2$ layers are non-magnetic \cite{eyert_metallic_2008,ong_origin_2010,daou_unconventional_2017}. The photoemission spectrum of PdCrO$_2$ contains prominent features, absent in PdCoO$_2$ \cite{noh_anisotropic_2009,noh_direct_2014,sunko_maximal_2017,sunko_probing_2020}, that can be understood from an effective inter-layer Kondo lattice model 
\cite{sunko_probing_2020}. The in-plane resistivities for the two compounds \cite{Hicks2015} are shown in Fig.~\ref{fig:model}(b), respectively. The salient features are as follows: (i) The magnitude of the resistivity for both compounds is small, suggesting that they are ``good" metals with a long mean-free path. (ii) PdCrO$_2$ is considerably more resistive than PdCoO$_2$ over a wide range of temperatures. (iii) PdCrO$_2$ displays a prominent $T-$linear scaling of the resistivity above $T\gtrsim 150$K (far above $T_{\rm N}\approx 37.5$K, the N\'eel temperature for $120^{\circ}-$antiferromagnetism) \cite{takatsu_critical_2009,takatsu_single_2010,takatsu_magnetic_2014,le_magnetic_2018}, and with a slope that is greater than the average slope of $\rho_{ab}(T)$ in PdCoO$_2$ in the same temperature range \cite{Supplementary_material}. 

 The central puzzle that we address in this Letter concerns the microscopic origin of the excess $T-$linear resistivity in PdCrO$_2$ relative to isostructural PdCoO$_2$ ($\Delta \rho \equiv \rho_{ab}^{\rm PdCrO_2} -\rho_{ab}^{\rm PdCoO_2}>0$), going beyond the conventional electron-phonon scattering mechanism, and in a temperature regime where the long-range magnetic order is lost. Given the contrast between PdCrO$_2$ and PdCoO$_2$, it is plausible that the fluctuations of the Cr-local moments play a crucial role on the electronic transport lifetimes even at the relatively high temperatures of interest (i.e. for $T\gtrsim T_N$). However, recent work \cite{mcroberts2022intermediatescale} has demonstrated that electrons scattering off the fluctuations of a ``cooperative" paramagnet \cite{Keren_1994,Moessner_1998L,Moessner_1998B,conlon,Anjana_2017,Mourigal_2019,Shu_2019,Knolle_2022T} can not account for a $T-$linear resistivity; instead the resistivity saturates to a temperature-independent value for $T\gg T_N$. Starting with a microscopic model, we will now demonstrate that the resolution to the conundrum lies in a previously ignored and non-trivial interaction term between the Pd-electrons, the Cr-spins, and phonons, as encoded in an electron-magneto-elastic (EME) coupling. Although specifically motivated by PdCrO$_2$, our theory has relevance beyond this single material. There are a number of exciting new material platforms that have come to the forefront in recent years that consist of stacks of metallic and Mott insulating layers \cite{moirerev,Mak2022,FaiMIT,pasupathyMIT,ajesh,ruhman,moireHF,vavno2021artificial,persky2022magnetic}. In what follows, we develop a general framework to address electrical transport in such layered material platforms, and our conclusions can be used to disentangle the various sources of interaction between electrons, local-moments and phonon degrees of freedom. 
 
 {\it Model.-} Consider a quasi two-dimensional (2D) layered model defined on a triangular lattice, where the electronic and local-moment degrees of freedom reside on alternating layers. The effective 2D Hamiltonian is given by (see \cite{Supplementary_material} for a microscopic derivation of $H_{\rm{EME}}$),
\begin{subequations}
\beq
  H &=& H_{\rm el} + H_{\rm S} + H_{\rm K} + H_{\rm ph} + H_{\rm el-ph} + H_{\rm EME},\nn \label{total_H} \\ \\
H_{\text{el}} & =&\sum_{\boldsymbol{k},\alpha}(\varepsilon_{\boldsymbol{k}} - \mu) p_{\boldsymbol{k}\alpha}^{\dagger}p^{\phantom\dagger}_{\boldsymbol{k}\alpha},\\
H_{\text{S}} & =&J_{\text{H}}\sum_{\left\langle i,j\right\rangle }\boldsymbol{S}_{i}\cdot\boldsymbol{S}_{j},\\
H_{\text{K}} & =&J_{\text{K}}\sum_{i} p^\dagger_{i\alpha} (\boldsymbol{S}_{i}\cdot\boldsymbol{\sigma}_{\alpha\beta}) p^{\phantom\dagger}_{i\beta}, \\
H_{\text{ph}} & =& \sum_{\ell=I_{\rm a},I_{\rm o},O}\sum_{\boldsymbol{q}} \left(\frac{|\pi^{(\ell)}_{\boldsymbol{q}}|^2}{2M} + \frac{M \omega_{\ell,\boldsymbol{q}}^2}{2} |\varphi^{(\ell)}_{\boldsymbol{q}}|^2 \right),\label{phonon_Hamil}\\
H_{\rm{el-ph}} & =& \sum_{i,\sigma} \bigg(\alpha_{\rm a} \nabla \varphi^{(I_{\rm a})}_{i}p_{{i}\sigma}^{\dagger}p^{\phantom\dagger}_{{i}\sigma}+\alpha_{\rm o} \varphi^{(I_{\rm o})}_{i}p_{{i}\sigma}^{\dagger}p^{\phantom\dagger}_{{i}\sigma}\bigg), \\
H_{\text{EME}} & =&\widetilde{\alpha}\sum_{i}\varphi^{(O)}_i p^\dagger_{i\alpha}  (\boldsymbol{S}_{i}\cdot\boldsymbol{\sigma}_{\alpha\beta})p^{\phantom\dagger}_{i\beta}.
\label{EME_Hamiltonian}
\eeq
\end{subequations} 
Here $p^{\dagger}_{\boldsymbol{k}\sigma},~p^{\phantom\dagger}_{\boldsymbol{k}\sigma}$ denote the Pd-electron creation and annihilation operators with momentum $\boldsymbol{k}$ and spin $\sigma=\pm 1/2$. The electronic dispersion is given by $\varepsilon_{\boldsymbol{k}}$ and $\mu$ is the chemical potential; experiments in PdCrO$_2$ indicate the Pd-electronic structure is well captured using first and second-neighbor hoppings and the conduction band, dominantly of Pd character, is very close to half-filling \cite{sunko_probing_2020, Hicks2015}. The local-moments, $\boldsymbol{S}_i$, interact mutually via nearest-neighbor antiferromagnetic Heisenberg exchange ($J_{\rm H}>0$), and with the Pd-electron spin-density via a Kondo exchange ($J_{\rm K}>0$), respectively. The Cr-electrons form $S=\frac{3}{2}$ local-moments \cite{sunko_probing_2020}. Finally, we include three phonon fields, $\varphi^{(\ell)}$ with $\ell=I_{\rm{a}},I_{\rm{o}},~O$, corresponding to in-plane acoustic ($I_{\rm a}$), in-plane optical ($I_{\rm o}$) and out-of-plane ($O$) lattice vibrations, respectively. We set the mass, $M$, to be equal for these modes for simplicity. The in-plane modes $I_{\rm a,o}$ couple to the $p$-electron density with strengths $\alpha_{\rm a,o}$, respectively, while the out-of-plane mode, $O$, couples to the ``inter-layer'' Kondo interaction with EME strength, $\widetilde{\alpha}$. We set the lattice constant $a=1$, unless stated otherwise. Importantly, we include couplings to both in-plane acoustic and optical modes to account for the full $T$-dependence of $\rho_{ab}$ in the absence of magnetism (i.e. in PdCoO$_2$) \cite{takatsu_roles_2007}. Henceforth, we neglect the weak momentum dependence and the form-factors associated with the different interaction terms in the quasi two-dimensional setting to simplify our discussion \cite{sunko_probing_2020}. We will restore these additional complexities when considering out-of-plane transport for reasons to be made clear below.
}

Let us begin by considering the simpler case where the optical modes are Einstein-phonons, with ${\omega}_{I_{\rm o},\boldsymbol{q}} = \omega_0$ and $\omega_{O,\boldsymbol{q}} = \widetilde{\omega}_0$, while for the acoustic mode ${\omega}_{I_{\rm a},\boldsymbol{q}} = cq$, with a corresponding Debye frequency $\omega_D$. The experimental 
regime of interest  corresponds to $\{\widetilde{\omega}_0,J_{\rm{H}}\}\lesssim T\ll\ve_F$, where $\ve_F$ is the Fermi energy for the Pd-electrons. Moreover, we shall consider the limit where $ J_{\rm{H}} \gg \{\widetilde{\alpha}\sqrt{\hbar/M\widetilde{\omega}_0},~J_{\rm{K}}\}$ and thereby ignore the feedback of both electrons and phonons on the properties of the local-moments. In recent work \cite{mcroberts2022intermediatescale}, some of us analyzed the properties of a subset of the terms, ($H_{\tn{el}}+H_{\tn{S}}+H_{\tn{K}}$), in Eq.~\eqref{total_H} at leading order in a small $J_{\rm{K}}$ by approximating the local-moments as $O(3)$ vectors, but capturing their complex precessional dynamics using the Landau-Lifshitz equations \cite{Keren_1994,Moessner_1998L,Moessner_1998B,conlon,Anjana_2017,Mourigal_2019,Shu_2019,Knolle_2022T}. While this leads to an interesting frequency dependence and momentum-dependent crossovers in the electronic self-energy, the temperature dependence can be understood entirely based on a high-temperature expansion with uncorrelated local-moments. The present manuscript will treat the local-moments on the same footing, but include the additional interaction effects due to ($H_{\rm{ph}}+H_{\rm{el-ph}}+H_{\rm{EME}}$).

{\it Results for intermediate-scale transport.-} We will analyze electrical transport for the model defined above within the framework of traditional Landau-Boltzmann paradigm \cite{Ziman_2001}. This is justified based on the magnitude of the resistivity being much smaller than the characteristic scale of $ha_B/e^2$ ($a_B\equiv$Bohr radius) and $k_F \ell_{\rm mfp} \gg 1$ over the entire temperature range of interest ($\ell_{\rm mfp}$ being the mean free path) \cite{Hicks2015}. Moreover, there is direct experimental evidence for the value of the dimensionless Kondo-coupling being small, based on recent photoemission experiments \cite{sunko_probing_2020}. 

\begin{figure}[t]
\centering
\includegraphics[width=0.45\textwidth]{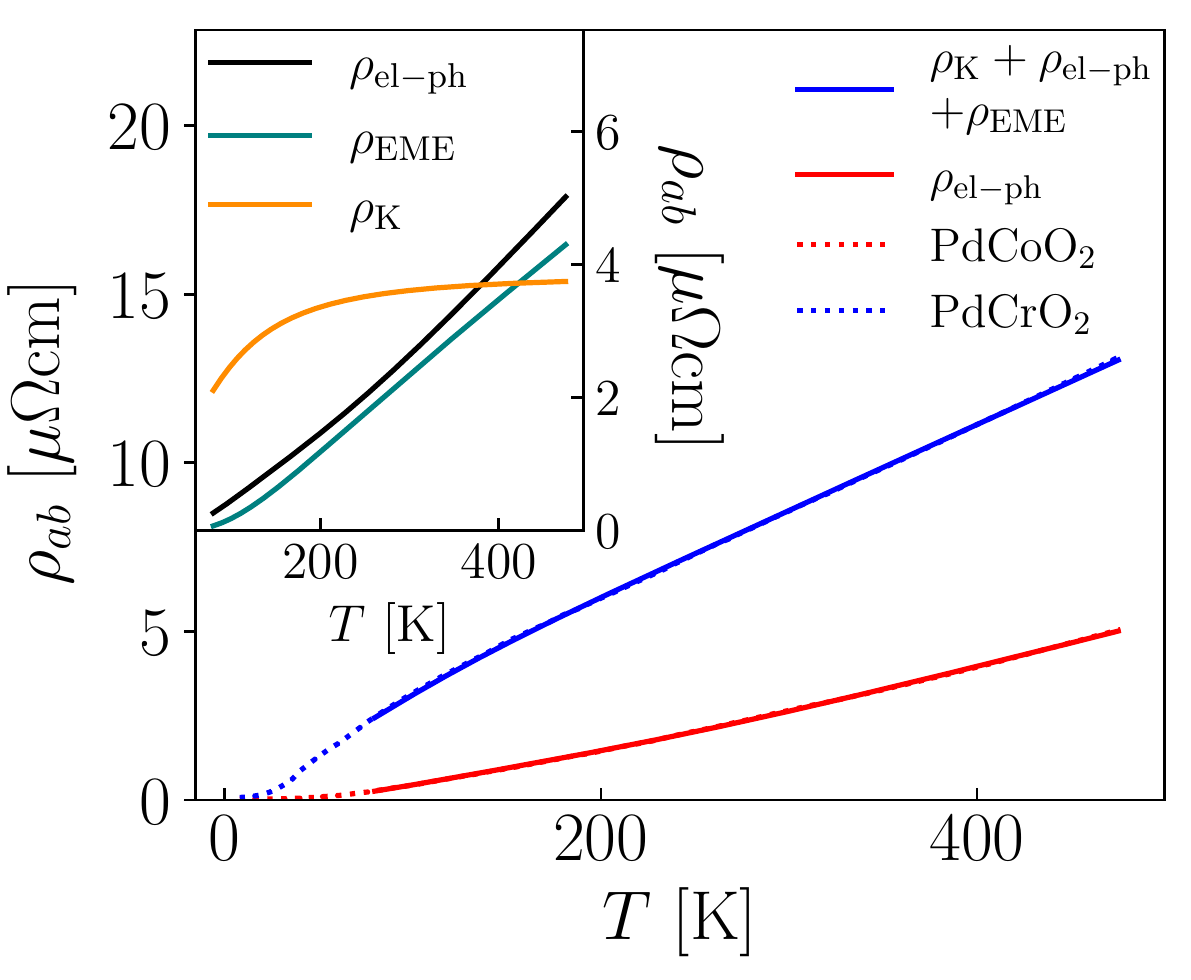} 
\caption{Comparison of $T-$dependence of in-plane resistivity between experiments \cite{Hicks2015} (dotted lines) and theoretical model in Eq.~\eqref{total_H} (solid lines). The only free parameters in the fits are the phonon frequencies and bare el-ph and EME couplings. All other parameters are fixed; see \cite{sunko_probing_2020,Supplementary_material}. For the phonon data, we have used $\omega_D = 29$meV, $\omega_0 = 120$meV, $\widetilde{\omega}_0 = 40$meV, $\lambda_{\rm{el-ph, a}} = 0.04$, $\lambda_{\rm{el-ph, o}} = 0.02$, $\lambda_{\rm EME} = 0.05$ and $S=3/2$.}
\label{fig:main}
\end{figure}

Considering the full $H$, we have multiple sources of scattering for the electrons. Within Boltzmann theory, the total transport scattering rate satisfies Matthiessen's rule \cite{Ziman_2001}: 
\begin{eqnarray}
\frac{1}{\tau_{\rm tr}} = \frac{1}{\tau_{\rm{el-ph}}} + \frac{1}{\tau_{\rm K}} + \frac{1}{\tau_{\rm EME}},
\label{tr_time}
\end{eqnarray}
and the in-plane resistivity is given by $\rho_{ab} = m/(ne^2 \tau_{\rm tr})$, where $m$ is the effective mass and $n$ is the electron density \cite{Supplementary_material}. Experiments with controlled amounts of irradiation shift the overall resistivity curves by a constant (and relatedly, the residual resistivity), without affecting the slope in the $T-$linear regime \cite{Zhakina2023}.  We start by describing the electron-phonon contribution, $1/\tau_{\rm{el-ph}}\equiv 1/\tau_{\rm{el-ph},a}+1/\tau_{\rm{el-ph},o}$.  
Previous works have obtained $1/\tau_{\rm{el-ph}}$ for PdCoO$_2$, and highlighted the importance of a high-frequency optical mode (which is not entirely in the equipartition regime at $T\lesssim \omega_0$) for the observed superlinear-scaling of $\rho_{ab}(T)$ \cite{takatsu_roles_2007,Hicks_mobility_2012}. We have fully reproduced these results for PdCoO$_2$ based on the same procedure \cite{Supplementary_material} within our 2D  model of the Fermi surface; see Fig.~\ref{fig:main}. In PdCrO$_2$, the scattering of electrons off the local-moments due to the bare Kondo interaction, $1/\tau_{\rm K}$, can lead to a sub-linear $T$-dependent contribution to $\rho_{ab}$ at temperatures $T \gtrsim J_{\rm H}$, before saturating to the $T$-independent value \cite{mcroberts2022intermediatescale} (see $\rho_{\rm K}=m/(ne^2 \tau_{\rm K})$ in the inset of Fig.~\ref{fig:main}).

We now turn to the important role of the EME term in PdCrO$_2$. For $\widetilde{\omega}_0\gtrsim J_{\rm{H}}$  we find that $\tau_{{\rm EME}}^{-1}$ follows closely the temperature dependence of the scattering rate of electrons interacting with an optical mode of frequency $\widetilde{\omega}_0$ with a modified dimensionless coupling, 
 \beq
\lambda_{\rm EME} = \frac{\nu_0 \widetilde{\alpha}^2}{{M}\widetilde{\omega}_0^2}S(S+1).
\label{lambdaeph}
\eeq
Here, $\nu_0$ is the electronic density of states at the Fermi level, and scattering off the (spin$-S$) local-moment fluctuations via the EME interaction leads to the additional factor of $[S(S+1)]$ \cite{Supplementary_material}. Ignoring the constant offset, $1/\tau_{\rm K}$ at $T\gg J_{\rm{H}}$, we find that
\begin{equation}
    \frac{1}{\tau_{\rm tr}} \approx \frac{2\pi}{\hbar}\left(\lambda_{\rm{el-ph, a}} +\lambda_{\rm{el-ph, o}} + \lambda_{\rm EME}\right) T,\quad T\gtrsim \omega_0,
    \label{slope_w_EME}
\end{equation}
where the dimensionless coefficients are given by,
\beq
\lambda_{\rm{el-ph,o}} = \frac{\nu_0\alpha^2_{\rm{o}}}{M\omega_0^2},~~\lambda_{\rm{el-ph,a}} = \frac{\nu_0\alpha^2_{\rm{a}}}{Mc^2}.
\label{lambdaeph}
\eeq
Consequently, at the highest temperatures, the effect of the EME term considered in this work is to enhance the slope ($A$) of a $T-$linear resistivity, $\rho_{ab}-\rho_0=AT$. This constitutes our first important result. 

Importantly, note that even if the bare EME coupling is weak relative to the electron-acoustic phonon coupling (i.e. $\widetilde{\alpha}/ \alpha_{\rm{a}} \ll 1$), the dimensionless coupling $\lambda_{\rm EME}$ is not necessarily small compared to $\lambda_{\rm{el-ph,a}}$. Furthermore, if the out-of-plane phonon 
is soft ($\widetilde{\omega}_0 / \omega_D \ll \widetilde{\alpha}/\alpha_{\rm{a}} $), the presence of the EME interaction can dramatically reduce the onset of $T$--linear resistivity to $\mathcal O(\widetilde{\omega}_0)$.

Assembling all of the above ingredients, we can now reproduce the resistivity in PdCrO$_2$, including the effect of the EME term; see Fig.~\ref{fig:main}. Note that the scattering rates due to the electron-phonon interaction in PdCoO$_2$ \cite{roles_highfreq_pdcoo2_takatsu_2007, Hicks_mobility_2012} and Kondo coupling in PdCrO$_2$ \cite{sunko_probing_2020} are fixed by previous experiments, which leaves two independent parameters in our theory --- $\lambda_{\rm EME}$ and $\widetilde{\omega}_0$ ($J_{\rm{H}}$ is also fixed \cite{sunko_probing_2020}). We determine these parameters by fitting the excess resistivity $\Delta\rho$ in Fig.~\ref{fig:model}(b) to the analytical form of $\tau_{{\rm EME}}^{-1}$ \cite{Supplementary_material}. The values obtained by this procedure are consistent with the characteristic out-of-plane lattice vibration frequency being naturally softer than the in-plane one,  $\omega_0\gtrsim\widetilde{\omega}_0\gtrsim J_{\rm H}$; however, we note that our theory extends beyond this regime. The resulting contribution to the resistivity, $\rho_{\rm EME}$, is shown in the inset of Fig.~\ref{fig:main}. Overall, the prominent $T$--linear resistivity at intermediate $T$ stems from (i) the EME scattering rate; and (ii) the combined sub- and super-linear contributions of $\rho_{\rm K}$ and $\rho_{\rm{el-ph}}$, respectively
 \cite{Supplementary_material}.  It is worth noting that the $T$-linear behavior in \eqref{slope_w_EME} applies when all phonon modes are in their equipartition regime. For PdCrO$_2$, this corresponds to $T\sim \mathcal{O}(1000)$K due to the large value of $\omega_0$, and is hence not directly related to the behavior presented in Fig.~\ref{fig:main}.

{\it Role of acoustic phonons.-} Our discussion thus far has focused on the simplified limit of an EME coupling to optical phonons. Let us now analyze the effects of an EME coupling to an acoustic phonon. For $H_{\text{ph}}$ in Eq.~\ref{phonon_Hamil}, this amounts to replacing $\widetilde{\omega}_0^2\left(\widetilde{\varphi}^{(O)}_{\vec{q}}\right)^{2}\rightarrow \widetilde{\omega}^2_{\boldsymbol{q}}\left(\widetilde{\varphi}^{(O)}_{\vec{q}}\right)^{2}$, where $\widetilde{\omega}_{\boldsymbol{q}}=\widetilde{c}q$ in the limit of small $q$. Similarly, for $H_{\text{EME}}$ in Eq.~\ref{EME_Hamiltonian}, this amounts to replacing $\varphi^{(O)}_i \to \nabla\varphi^{(O)}_i$. Note that, in practice, the out-of-plane vibrations that couple the layers are at a finite wavevector, namely, 
$\omega_{\boldsymbol{q}} \approx \sqrt{\widetilde{c}^2q_x^2+\widetilde{c}^2q_y^2+ \widetilde{c}_z^2q_{z,0}^2}$ in the limit where $\widetilde{c}_z q_{z,0}\ll T$.  
It is worth noting that unlike the conventional electron-phonon interaction, where scattering is mainly small-angle up to the BG temperature, the EME term induces large-angle scattering even at low-$T$ due to large momentum transfer to the local-moments, which serve as a ``bath". If the spin structure-factor exhibits non-trivial correlations in the Brillouin-zone (i.e. the spin-correlation length is finite with remnants of Bragg-like peaks), the intermediate-scale transport behavior is controlled by the Pd-electron Fermi-surface geometry. However, when the spin-correlation length is short, $1/\tau_{\rm EME}$ shows two distinct regimes~\cite{Supplementary_material}. For $T \gtrsim \widetilde{T}_{\rm BG}\equiv 2\widetilde{c}k_F$, the Bloch-Gr\"uneisen temperature, the result reduces to the case of optical phonons, $1/\tau_{\rm EME}=2\pi \lambda_{\rm EME}T$ with $\lambda_{\rm EME} = \frac{\nu_0 \widetilde{\alpha}^2}{M\widetilde{c}^2}S(S+1)$. On the other hand, for $J_{\rm H} \lesssim T \lesssim \widetilde{T}_{\rm BG}$, we encounter an unexpected $1/\tau_{\rm EME}=2\pi \lambda_{\rm EME}T^2/\widetilde{T}_{\rm BG}$, instead of the usual $\sim T^4$ regime in two-dimensions for the phase-space reasons introduced above. Interestingly, this is an example of a $T^2$ ``quasielastic" scattering due to the EME term (instead of the usual $T^2$ due to umklapp scattering).

\begin{figure}[t]
\centering
\includegraphics[width=0.45\textwidth]{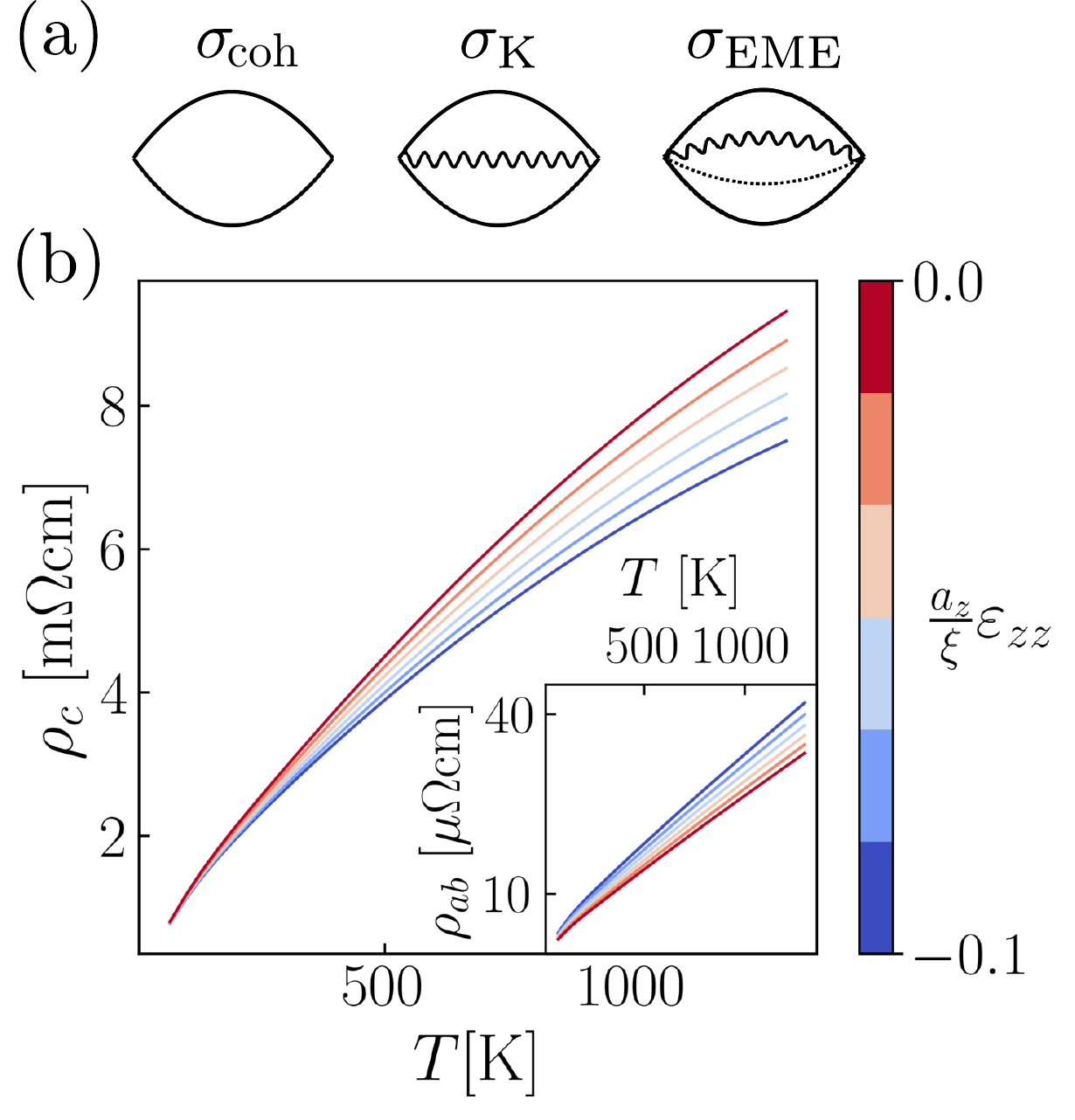} 
\caption{(a) The Feynman diagrams contributing at leading-order to the $c-$axis conductivity, $\sigma_c$ (see Eq.~\eqref{c_axis_cond_terms}), where solid, wiggly and dashed lines denote electrons, spins and phonons, respectively. (b) $c$-axis resistivity as a function of $T$ for different (compressive) $c$-axis strain $\varepsilon_{zz}$ (in units of $\xi/a_z$, where $\xi$ is the characteristic decay length of the hopping integral and $a_z$ is the $\hat{z}$ lattice constant). Inset: in-plane resistivity as a function of $T$ under $c$-axis strain $\varepsilon_{zz}$. 
}
\label{fig:predictions}
\end{figure}

\textit{Effect of $c$-axis strain on in-plane transport.-} Given that the proposed EME interaction in PdCrO$_2$ originates from fluctuations of the (inter-layer) Kondo coupling, applying $c$-axis pressure is expected to enhance the slope of the $T$--linear resistivity for the following reason. The bare EME coupling is controlled in part by the Kondo-scale, $\widetilde{\alpha}\propto t_{cp}^2/U$ where $t_{cp}$ is the inter-plane hybridization between the Pd and Cr-electrons and $U$ is the on-site Coulomb repulsion for Cr-electrons \cite{sunko_probing_2020,Supplementary_material}. Upon applying $c$-axis strain, the inter-layer distance reduces, thereby increasing $t_{cp}$, which is exponentially sensitive to the deformation; the stiffening of the out-of-plane phonons is at best algebraic. Therefore, the dimensionless EME coupling, $\lambda_{\rm EME}\propto t^4_{cp}/(U\widetilde{\omega}_0)^2$ is expected to show a significant increase, along with an enhancement of the Kondo coupling which affects the constant shift in the resistivity at high $T$. The predicted form of the in-plane transport is depicted in the inset of Fig.~\ref{fig:predictions}(b) for a range of $c-$axis strain ($\varepsilon_{zz}$) for bare microscopic parameters as chosen in Fig.~\ref{fig:main}.

\textit{Out-of-plane transport.-} The electrical resistivity along the $c$-axis provides a direct window into the inter-layer nature of the magnetic interactions, which we have absorbed so far in the effective 2D model. We consider the leading contributions to the $c$-axis conductivity within linear-response theory, which is given by 
\begin{eqnarray}
\sigma_c = \sigma_{\rm coh} + \sigma_{\rm K} + \sigma_{\rm EME},
\label{c_axis_cond_terms}
\end{eqnarray}
where $\sigma_{\rm coh}$ arises from the ``coherent'' channel due to inter-layer $p$ to $p$ hoppings ($t_{pp}$), and $\sigma_{\rm K}$, $\sigma_{\rm EME}$ represent the ``incoherent'' channels due to inter-layer spin-assisted and spin-phonon-assisted contributions, respectively \cite{Supplementary_material}. For simplicity, we ignore the contribution due to an interlayer ``incoherent'' phonon-assisted hopping not involving the local moments; such a term does not affect our results at a qualitative level. The leading-order Feynman diagrams corresponding to each of these contributions  are depicted in Fig.~\ref{fig:predictions}(a). We have $\sigma_{\rm coh} = e^2 (n/m)_c \tau_{\rm tr}$, where $(n/m)_c$ is related to the $c$-axis dispersion and $\tau_{\rm tr}$ is given by Eq.~\eqref{tr_time}; the detailed expressions for the incoherent channels appear in \cite{Supplementary_material}.

The coherent channel dominates $\sigma_c$ up to temperatures $T\lesssim T_* \sim 10^3~$K (determined by the condition $\sigma_{\rm coh}(T_*) \approx \sigma_{\rm EME}(T_*)$ \cite{Supplementary_material}), such that, in this $T$-regime, $\sigma_c$ and $\sigma_{ab}$ follow the same $T$-scaling as they share the same transport lifetime. The contribution of the incoherent channels becomes significant at temperatures $T\gtrsim T_*$, where, due to the weak $c$-axis dispersion, their temperature dependence is determined by the $T$-scaling of the current vertices, rather than the transport lifetime. In particular, for $T\gtrsim \omega_D,\widetilde{\omega}_0$, while $\sigma_{\rm coh} \sim 1/T$ as in the in-plane case, $\sigma_{\rm K}$ is independent of $T$ and $\sigma_{\rm EME}\propto T$. As a result, the $c$-axis resistivity becomes sublinear at sufficiently high temperatures, as depicted in Fig.~\ref{fig:predictions}(b).

The interplay between the different conduction mechanisms has signatures in the behavior under c-axis pressure. The in-plane conductivity is expected to decrease with $c-$axis compression, since the EME scattering is enhanced (because it is proportional to the inter-plane hopping strength.)
In contrast, the coherent part of the $c-$axis conductivity increases, as a result of the enhanced inter-layer hopping~\cite{Supplementary_material}. Interestingly, this increase in $\sigma_{\rm coh}$ is associated solely with the el--ph scattering term. The contributions to $\sigma_{\rm coh}$ from the Kondo and EME terms are proportional to $t_{pp}^2/t_{cp}^4$~\cite{Supplementary_material}; assuming that $t_{pp}\propto t_{cp}^2$,  
the ratio $t_{pp}^2/t_{cp}^4$ is unchanged by $c-$axis strain. Similarly, the incoherent parts of the conductivity increase with compression. However, they do so slightly in excess of $\sigma_{\rm coh}$ which in turn reduces the crossover scale $T_*$, as manifested by the increase in curvature at intermediate temperatures with increasing compression; see Fig.~\ref{fig:predictions}(b) \cite{Supplementary_material}.  There are measurements of the $c-$axis resistivity in the literature \cite{Takatsu_2010,Ghannadzadeh2017}, but the reported values are inconsistent with each other. The reason for this discrepancy is currently unclear. To resolve these issues, more accurate measurements are needed, using e.g. the techniques described in Ref.~\cite{putzke}.

{\it Discussion.-} Our conjectured magneto-elastic mechanism for  the enhanced $T-$linear resistivity in PdCrO$_2$ relies on quasi-elastic scattering, where the phonons are in the equipartition regime. There is experimental evidence for the Lorenz ratio satisfying the Wiedemann-Franz law in the $T-$linear regime \cite{Zhakina2023}, which is consistent with our mechanism. Interestingly, the extracted transport scattering rate in the same regime of $T-$linear resistivity is Planckian with $C\approx 0.9$ \cite{Zhakina2023}. Within our model and in the quasi-elastic regime, this is not indicative of any fundamental principle, such as a bound associated with an inelastic scattering rate. It is, however, far from obvious why the scattering rate turns out to be Planckian. 

A natural future direction is to study the effects of the EME term at low temperature.  In particular, the magneto-elastic coupling might be evident in the electronic spectral function. Signatures of phonon drag observed in PdCoO$_2$ \cite{Hicks_mobility_2012} are expected to be suppressed in PdCrO$_2$ due to EME-induced large-angle scattering of phonons off magnetic moments \cite{takatsu_critical_2009}. Pronounced magnetic correlations may also lead to a generalized Kohn anomaly \cite{Kohn_anomaly} associated with phonon softening at the AFM wavevectors. A detailed understanding of the low-temperature properties of this interesting system remains an open problem.

{\it Acknowledgements-} JFMV and DC acknowledge the hospitality of the Weizmann Institute of Science, where this work was completed. JFMV and DC thank A. McRoberts and R. Moessner for an earlier related collaboration \cite{mcroberts2022intermediatescale} and insightful discussions. DC thanks P. Coleman, J. Ruhman, and T. Senthil for useful discussions. JFMV and DC are supported in part by a CAREER grant from the NSF to DC (DMR-2237522). DC and EB acknowledge the support provided by the Aspen Center for Physics where this collaboration was initiated, which is supported by National Science Foundation grant PHY-1607611. Research in Dresden benefits from the environment provided by the DFG Cluster of Excellence ct.qmat (EXC 2147, project ID 390858940).

\bibliography{refs.bib}

\clearpage
\renewcommand{\thefigure}{S\arabic{figure}}
\renewcommand{\figurename}{Supplemental Figure}
\setcounter{figure}{0}
\appendix
\pagenumbering{arabic}
\begin{widetext}

\begin{center}
  \textbf{\large SUPPLEMENTARY INFORMATION}\\[.2cm]
  \textbf{\large $T-$linear resistivity from magneto-elastic scattering: application to PdCrO$_2$}\\[.2cm]
  J.F. Mendez-Valderrama$^{1,*}$, Evyatar Tulipman$^{2,*}$, Elina Zhakina$^{3}$, \\ Andrew P. Mackenzie$^{3,4}$, Erez Berg$^{2}$, and Debanjan Chowdhury$^{1}$
  
  {\itshape
        \mbox{$^{1}$Department of Physics, Cornell University, Ithaca, New York 14853, USA.}\\
        \mbox{$^{2}$Department of Condensed Matter Physics, Weizmann Institute of Science, Rehovot 76100, Israel}\\
  	\mbox{$^{3}$Max Planck Institute for Chemical Physics of Solids, N\"othnitzer Stra\ss e 40, 01187 Dresden, Germany}\\
   \mbox{$^{4}$Scottish Universities Physics Alliance, School of Physics \& Astronomy, }
   \mbox{University of St. Andrews, St. Andrews KY16 9SS, United Kingdom}
	
 }
\end{center}
\let\thefootnote\relax\footnote{* These authors contributed equally to this work}

\section{Derivation of the EME coupling} \label{appendix_model}

We derive the low-energy effective Kondo lattice model with EME term starting from an Anderson-type model coupled to phonons; the Kondo term was already derived in Ref.~\cite{sunko_probing_2020}.
The key new insight is as follows: the effective Kondo-like coupling
that is generated between the Pd-electrons and Cr local moments involves
an inter-layer hybridization, $t_{cp,ij}$ between sites $i$ and $j$.
If this coupling were to be affected by the fluctuations associated
with the corresponding out-of-plane bond, $t_{cp,ij}$ will be renormalized
by the displacement, $\varphi^{(O)}_{ij}$. Starting with the microscopic
Hamiltonian,
\begin{align}
H & =H_{\text{Pd}}+H_{\text{Cr}}+H_{\text{hyb}}+H_{\text{ph,EME}},\\
H_{\text{Pd}} & =\sum_{\boldsymbol{k}\sigma}\left(\varepsilon_{\boldsymbol{k}}-\mu\right)p_{\boldsymbol{k}\sigma}^{\dagger}p_{\boldsymbol{k}\sigma},\\
H_{\text{Cr}} & =-\sum_{ij\sigma}t_{ij}^{c}c_{i}^{\dagger}c_{j}+\text{h.c}+U\sum_{i}\left(n_{i\uparrow}^{c}-\frac{1}{2}\right)\left(n_{i\downarrow}^{c}-\frac{1}{2}\right),\\
H_{\text{hyb}} & =\sum_{\left\langle ij\right\rangle \sigma}t_{cp,ij}\left(1-\eta\varphi^{(O)}_{ij}\right)\left(c_{i\sigma}^{\dagger}p_{j\sigma}+\text{h.c}\right)\\
H_{\text{ph,EME}} & =\sum_{\left\langle ij\right\rangle }\frac{\pi_{ij}^{(O)\,2}}{2M}+\frac{M\tilde{\omega}_{0}}{2}\varphi_{ij}^{(O)\,2},
\end{align}
where $c,~c^\dagger$ represent the Cr-electron operators, $t_{ij}^{c}$ is a nearest-neighbor hopping in the Cr layer, $\varepsilon_{\boldsymbol{k}}$
is the dispersion for the Pd electrons, $\eta$ determines the scale by which
the hybridization is renormalized to leading order in the bond displacement
$\varphi^{(O)}_{ij}$, $M$ is the phonon mass, $n_{i\sigma}^{c}$ is the
density operator for Cr electrons, and $U$ is the on site interaction
strength. In general, these interactions  involve multiple orbitals associated with the Cr crystal field multiplets, and contain the full lattice structure of PdCrO$_{2}$. However, here for simplicity, we only retain a single orbital per site in the Cr layers.

We consider the strong-coupling limit, $U\gg t^{c},g$ (similarly to Ref.~\cite{sunko_probing_2020}), and derive the low-energy model based on a Schriffer-Wolff transformation. Retaining terms up to linear-order in the bond displacement, we obtain:
\begin{align}
H & =H_{\text{Pd}}+H_{\text{lm}}+H_{\text{K}}+H_{\text{EME}}+H_{\text{ph,EME}},\\
H_{\text{lm}} & =J_{H}\sum_{\left\langle ij\right\rangle }\boldsymbol{S}_{i}\cdot\boldsymbol{S}_{j}\\
H_{\text{K}} & =\sum_{\left\langle ilj\right\rangle \sigma}\frac{4t_{cp,ij}t_{cp,lj}}{U}\boldsymbol{S}_{l}\cdot\left(p_{i}^{\dagger}\boldsymbol{\sigma}p_{j}\right)\\
H_{\text{EME}} & =-\sum_{\left\langle ilj\right\rangle \sigma}\frac{4t_{cp,ij}t_{cp,lj}}{U}\eta\left(\varphi^{(O)}_{il}+\varphi^{(O)}_{jl}\right)\boldsymbol{S}_{l}\cdot\left(p_{i}^{\dagger}\boldsymbol{\sigma}p_{j}\right)
\end{align}

The first term above is the usual Heisenberg coupling with antiferromagnetic
exchange $J_{H}=4(t^{c})^2/U$. The second term  is a Kondo coupling \cite{sunko_probing_2020} with the appropriate form-factors coming from the fact that the interaction is inter-layer. The third term is the new EME coupling
between local moments in the Cr layer, optical phonons and itinerant Pd-electrons. In what follows, we consider the simplified case where the adjacent Pd layers are stacked in a perfectly aligned arrangement, and the hybridization is independent of the bond direction, i.e. $t_{cp,ij}\rightarrow t_{cp}$. In
this limit, we can define the Kondo and EME couplings to be $J_{K}=4t_{cp}^{2}/U$
and $\tilde{\alpha}=-J_{K}\eta$. We then have the following Hamiltonian
in momentum space
\begin{equation}
H=H_{\text{Pd}}+H_{\text{ph}}+H_{\text{lm}}+H_{\text{int}}
\end{equation}

with the interaction given by
\begin{align}
H_{\text{int}}= & \frac{1}{\sqrt{N}}\sum_{\substack{\bm{k}\bm{k}'\\
\alpha\beta
}
}J_{K}\left(\boldsymbol{k},\boldsymbol{k}'\right)p_{\boldsymbol{k}\alpha}^{\dagger}\left[\boldsymbol{S}_{\boldsymbol{k}-\boldsymbol{k}'}\cdot\boldsymbol{\sigma}_{\alpha\beta}\right]p_{\boldsymbol{k}'\beta}\\
 & +\frac{1}{N}\sum_{\bm{k}\bm{k}'\bm{q}}\sum_{\ell\alpha\beta}\tilde{\alpha}_{\ell}\left(\boldsymbol{k},\boldsymbol{k}'\right)\varphi^{(O)}_{\ell}\left(\boldsymbol{k}-\boldsymbol{k}'-\boldsymbol{q}\right)p_{\boldsymbol{k}\alpha}^{\dagger}\left[\boldsymbol{S}_{\boldsymbol{q}}\cdot\boldsymbol{\sigma}_{\alpha\beta}\right]p_{\boldsymbol{k}'\beta}
\end{align}

where the Fourier transformed operators are:
\begin{equation}
\varphi^{(O)}_{\ell+j,j}=\frac{1}{\sqrt{N}}\sum_{\bm{k}}e^{-i\bm{k}\bm{R}_{j}}\varphi^{(O)}_{\ell}\left(\boldsymbol{k}\right)
\end{equation}

\begin{equation}
p_{i\alpha}^{\dagger}=\frac{1}{\sqrt{N}}\sum_{\bm{k}}e^{-i\bm{k}\bm{R}_{i}}p_{\boldsymbol{k}\alpha}^{\dagger}
\end{equation}
\begin{equation}
\boldsymbol{S}_{j}=\frac{1}{\sqrt{N}}\sum_{\bm{q}}e^{-i\bm{q}\bm{R}_{j}}\boldsymbol{S}_{\boldsymbol{q}}
\end{equation}

and the momentum-dependent couplings are 
\begin{align*}
J_{K}\left(\boldsymbol{k},\boldsymbol{k}'\right) & =J_{K}\left[2\cos\left(k_{z}a_{z}\right)\right]\left[2\cos\left(k_{z}'a_{z}\right)\right]\\
\tilde{\alpha}_{+1}\left(\boldsymbol{k},\boldsymbol{k}'\right) & =\tilde{\alpha}_{-1}\left(\boldsymbol{k},\boldsymbol{k}'\right)=\tilde{\alpha}\left(\boldsymbol{k},\boldsymbol{k}'\right)=-\eta J_{K}\left(\boldsymbol{k},\boldsymbol{k}'\right),
\end{align*}
with $\bm{R}_{i}$ being the position of site $i$ and $N$ the number
of unit cells. 

We have introduced three new parameters: the mass of the out-of-plane mode $M$ (appearing in the EME term), its frequency
$\tilde{\omega}_{0}$ and the EME coupling $\eta$. As we detail
in the text, these parameters only enter the theory either in the
precise combination that corresponds to the dimensionless EME coupling,
or in the combination $\beta\tilde{\omega}_{0}$, leaving effectively
two new parameters in the theory. The rest of the parameters are taken
from the values provided in \cite{sunko_probing_2020}. Specifically, we have $U=4$eV
and $t_{cp}=107$meV which lead to a Kondo coupling of $J_{K}=11.45$ meV.
Additionally, we have $J_{H}=6.25$meV, $t_{nn}^{p}=568$meV, $t_{nnn}^{p}=108$meV,
and $\mu=256$meV. We fit the difference
of the resistivity of PdCrO$_{2}$ and PdCoO$_{2}$ to obtain the
dimensionless EME coupling in the main text and the frequency of this mode. 

\section{Transport rate} \label{transport_rate}

Within the variational Boltzmann approach \cite{Ziman_2001} 
the transport
lifetime can estimated by 
\begin{equation}
\rho\leq\frac{\frac{V}{2k_BT}\sum_{\boldsymbol{q}\boldsymbol{p}}P_{\boldsymbol{p}\boldsymbol{q}}\left(\psi_{\boldsymbol{q}}-\psi_{\boldsymbol{p}}\right)^{2}}{\left|\sum_{\boldsymbol{q}}e\frac{\partial f}{\partial\epsilon_{\boldsymbol{q}}}\psi_{\boldsymbol{q}}\boldsymbol{v}\left(\boldsymbol{q}\right)\cdot\boldsymbol{n}\right|^{2}}
\end{equation}
with $\psi_{\boldsymbol{q}}$ a variational anzatz, $P_{\boldsymbol{p}\boldsymbol{q}}$
the transition probaility from different momentum states, $V$ the
volume, $\boldsymbol{n}$ is a unit vector in the direction of the electric field $\boldsymbol{E}$,
$f$ the Fermi-Dirac distribution, $e$ the electron charge, $\boldsymbol{v}\left(\boldsymbol{q}\right)=\partial_{\boldsymbol{q}}\epsilon_{\boldsymbol{q}},$
and $\epsilon_{\boldsymbol{q}}$ the dispersion. Using the ansatz
$\psi_{\boldsymbol{q}}=-e\boldsymbol{v}_{\boldsymbol{q}}\cdot\boldsymbol{E}\tau_{\text{tr}}$
we recover the drude formula with the momentum relaxation time
\begin{equation}
\tau_{\text{tr}}^{-1}=\frac{\frac{1}{k_BT}\sum_{\boldsymbol{q}\boldsymbol{p}}P_{\boldsymbol{p}\boldsymbol{q}}\left(\boldsymbol{v}\left(\boldsymbol{q}\right)\cdot\boldsymbol{n}-\boldsymbol{v}\left(\boldsymbol{q}\right)\cdot\boldsymbol{n}\right)^{2}}{\sum_{\boldsymbol{q}}2\left(-\frac{\partial f}{\partial\epsilon_{\boldsymbol{q}}}\right)\left(\boldsymbol{v}\left(\boldsymbol{q}\right)\cdot\boldsymbol{n}\right)^{2}}.
\label{tr_time_SI}   
\end{equation}

At intermediate-$T$, where the scattering mechanisms are essentially isotropic, Mathiessen's rule is obeyed such that 
\begin{equation}
\tau_{\text{tr}}^{-1}=\tau_{\text{K}}^{-1}+\tau_{\text{EME}}^{-1}+\tau_{\text{el--ph}}^{-1}
\end{equation}
where each individual scattering rate is determined by the transition
probabilities 
\begin{equation}
P_{\boldsymbol{p}\boldsymbol{q}}^{\rm{K}}=k_BT\frac{2\pi}{\hbar}\left(-\frac{\partial f_{\boldsymbol{p}}}{\partial\epsilon_{\boldsymbol{p}}}\right)\left|J_{K}\left(\boldsymbol{p},\boldsymbol{q}\right)\right|^{2}\chi''\left(\boldsymbol{p}-\boldsymbol{q},\varepsilon_{\boldsymbol{p}}-\epsilon_{\boldsymbol{q}}\right)\left(n_{B}\left(\varepsilon_{\boldsymbol{p}}-\epsilon_{\boldsymbol{q}}\right)+n_{F}\left(-\epsilon_{\boldsymbol{q}}\right)\right),
\end{equation}

\begin{align}
P_{\boldsymbol{k}\boldsymbol{p}}^{\rm{EME}} =k_BT\frac{2\pi}{\hbar}\left(-\frac{\partial f}{\partial\epsilon_{\boldsymbol{p}}}\right)\left|\tilde{\alpha}\left(\boldsymbol{p},\boldsymbol{q}\right)\right|^{2} & \sum_{\boldsymbol{k}}\int\frac{d\nu}{\pi}\chi''\left(\boldsymbol{k}+\boldsymbol{p}-\boldsymbol{q},\epsilon_{\boldsymbol{p}}-\nu\right)\mathcal{B}\left(\boldsymbol{k},\nu-\varepsilon_{\boldsymbol{q}}\right) \times \nonumber \\ &\left[n_{B}\left(\epsilon_{\boldsymbol{p}}-\nu\right)+n_{F}\left(-\nu\right)\right]\left[n_{B}\left(\nu-\varepsilon_{\boldsymbol{q}}\right)+n_{F}\left(-\varepsilon_{\boldsymbol{q}}\right)\right],
\end{align}

\begin{equation}
P_{\boldsymbol{p}\boldsymbol{q}}^{\rm{el-ph}}=kT\frac{2\pi}{\hbar}\left(-\frac{\partial f_{\boldsymbol{p}}}{\partial\epsilon_{\boldsymbol{p}}}\right)\sum_{\lambda}\left|\alpha_{\lambda}\left(\boldsymbol{p},\boldsymbol{q}\right)\right|^{2}\mathcal{B}_{\lambda}\left(\boldsymbol{p}-\boldsymbol{q},\varepsilon_{\boldsymbol{p}}-\epsilon_{\boldsymbol{q}}\right)\left(n_{B}\left(\varepsilon_{\boldsymbol{p}}-\epsilon_{\boldsymbol{q}}\right)+f\left(-\epsilon_{\boldsymbol{q}}\right)\right).
\end{equation}
Here $\chi''\left(\boldsymbol{p},\omega\right)$ is the spin susceptibility
and $\mathcal{B}_{\lambda}\left(\boldsymbol{k},\nu\right)$ is the
phonon spectral function for mode $\lambda=o,a$ which can be either
acoustic or optical. We also reintroduced the momentum dependence
in all the couplings. In the case where the magnetic and itinerant
electron layers are perfectly aligned we have 
\begin{equation}
J_{K}\left(\boldsymbol{p},\boldsymbol{q}\right)=J_{K}\left[2\cos\left(p_{z}a_{z}\right)\right]\left[2\cos\left(q_{z}a_{z}\right)\right],
\end{equation}

\begin{equation}
\tilde{\alpha}\left(\boldsymbol{p},\boldsymbol{q}\right)=\tilde{\alpha}\left[2\cos\left(p_{z}a_{z}\right)\right]\left[2\cos\left(q_{z}a_{z}\right)\right],
\end{equation}

\begin{equation}
\alpha_{\rm{o}}\left(\boldsymbol{p},\boldsymbol{q}\right)=\alpha_{\rm{o}},
\end{equation}

and 

\begin{equation}
\alpha_{\rm{a}}\left(\boldsymbol{p},\boldsymbol{q}\right)=\alpha_{\rm{a}}\left|\boldsymbol{p}-\boldsymbol{q}\right|.
\end{equation}
Here the coupling constants are as defined in the main text. Additionally,
we make the local approximation for the spins \cite{mcroberts2022intermediatescale}:
\begin{equation}
\chi''\left(\nu\right)=S\left(S+1\right)\beta\nu\frac{\theta\left(2\pi J_{H}-\left|\nu\right|\right)}{2J_{H}}
\end{equation}
with the sum rule constraint 
\begin{equation}
\int\frac{\chi''\left(\nu\right)}{\beta\nu}d\nu=2\pi S\left(S+1\right).
\end{equation}
The spectral function for the free phonons is given by 
\begin{equation}
\mathcal{B}_{\lambda}\left(\boldsymbol{k},\nu\right)=\frac{\pi\hbar}{2M\omega_{\boldsymbol{k},\lambda}}\left(\delta\left(\nu-\omega_{\boldsymbol{k},\lambda}\right)-\delta\left(\nu+\omega_{\boldsymbol{k},\lambda}\right)\right)
\end{equation}
where for optical phonons we have $\omega_{\boldsymbol{k},o}=\omega_{0}$
while for acoustic phonons $\omega_{\boldsymbol{k},a}=ck$. The integrals
above can then be evaluated analytically if we take a quasi-2D circular
fermi surface. This leads to the total scattering rate: 
\begin{align}
\tau_{\rm{tr}}^{-1} & =2\pi\frac{\lambda_{tr,\rm{EP},o}}{\hbar\beta}\mathcal{I}_{1}\left(\frac{\beta\hbar\omega_{0}}{2}\right)+2\pi\frac{\lambda_{tr,\rm{EP},a}}{\hbar\beta}\mathcal{I}_{2}\left(\frac{\beta\hbar\omega_{BG}}{2}\right) \nonumber\\
 & +2\pi\frac{\lambda_{K}J_{\rm{K}}}{\hbar}\mathcal{I}_{3}\left(2\pi\beta J_{\rm{H}}\right)+2\pi\frac{\lambda_{tr,\rm{EME}}}{\hbar\beta}\mathcal{I}_{4}\left(2\pi\beta J_{\rm{H}},\beta\hbar\tilde{\omega}_{0}\right)
 \label{eqn:transport_rate}
\end{align}

Here we define
\begin{equation}
\mathcal{I}_{1}\left(x\right)=\left[x\text{csch}\left(x\right)\right]^{2},
\label{def_of_K}
\end{equation}
and
\begin{equation}
\mathcal{I}_{2}\left(x\right)=\frac{4}{\pi}\frac{1}{x^{3}}\int_{0}^{x}dy\ \frac{y^{4}}{\sqrt{1-\left(\frac{y}{x}\right)^{2}}}\left[\text{csch}\left(y\right)\right]^{2},
\end{equation}
corresponding to the case of acoustic phonons
interacting with a circular
Fermi surface in two dimensions. In addition,
\begin{align}
\mathcal{I}_{3}\left(y\right) & =-\frac{4\text{Li}_{2}\left(e^{-y}\right)}{y}-\frac{2y}{e^{y}-1}-4y+\frac{2\pi^{2}}{3y}+4\log\left(e^{y}-1\right),
\end{align}
corresponds to the pure Kondo scattering at high temperatures, where the logarithmic integrals result from integrating the Fermi function.
And finally 
\begin{equation}
\mathcal{I}_{4}\left(y,z\right)=\frac{z^{2}}{\cosh\left(z\right)-1}\mathcal{J}\left(y,z\right)
\end{equation}
is the result of the integrals involving the EME coupling, with the
additional definition:
\begin{align}
\mathcal{J}\left(y,z\right) & =\frac{-2z\text{Li}_{2}\left(e^{-y}\right)+\left(2y+z\right)\text{Li}_{2}\left(e^{-y-z}\right)+\left(z-2y\right)\text{Li}_{2}\left(e^{z-y}\right)+2\text{Li}_{3}\left(e^{-y-z}\right)-2\text{Li}_{3}\left(e^{z-y}\right)}{2yz} \nonumber \\
 & +\frac{z^{2}}{12y}-\frac{1}{2}\log\left(\left(e^{y}-e^{z}\right)\left(e^{y+z}-1\right)\right)+\frac{y\log\left(\sinh\left(\frac{y-z}{2}\right)\text{csch}\left(\frac{y+z}{2}\right)\right)}{2z}+\frac{y}{2}+\frac{\pi^{2}}{3y}+\log\left(e^{y}-1\right)+\frac{z}{2}.
\end{align}

These functions have the following high temperature limits which can
be obtained from the asymptotic expansions for the polylogarithm
\begin{equation}
\lim_{T\rightarrow\infty}\mathcal{I}_{1}\left(\frac{\beta\hbar\omega_{0}}{2}\right)=\lim_{T\rightarrow\infty}\mathcal{I}_{2}\left(\frac{\beta\hbar\omega_{BG}}{2}\right)=1
\end{equation}

and 
\begin{equation}
\lim_{T\rightarrow\infty}\mathcal{I}_{3}\left(2\pi\beta J_{H}\right)=\lim_{T\rightarrow\infty}\mathcal{I}_{4}\left(2\pi\beta J_{H},\beta\hbar\tilde{\omega}_{0}\right)=2.
\end{equation}

In the main text, we use the effective mass of PdCoO$_{2}$ extracted
from quantum oscillations given in Ref.~\cite{sunko_angle_2019}, and the density corresponding
to half filling to extract the transport rate from experiments following
the procedure in Ref.~\cite{bruin_similarity_2013} to calculate the ratio $n/m$ for both PdCoO$_{2}$ and PdCrO$_{2}$. Then we fit the resulting scattering
rate to the first two terms which correspond to electron-phonon scattering
in $\tau_{tr}^{-1}$. With this procedure, we find the parameters
$\lambda_{\rm el-ph,o}=0.024$,$\hbar\omega_{0}=120$meV,$\lambda_{\rm el-ph,a}=0.043$,
and $\hbar\omega_{\rm BG}=29$meV, which are in close agreement with the
reported values in \cite{Hicks_mobility_2012, roles_highfreq_pdcoo2_takatsu_2007}. The difference between our treatment
and that in Refs.~\cite{Hicks_mobility_2012, roles_highfreq_pdcoo2_takatsu_2007} is that we consider a 2D fermi surface instead
of a 3D Fermi surface. At intermediate temperatures
where our theory is valid we do not see significant differences in
the fit quality between the model considered here and the ones in
\cite{Hicks_mobility_2012, roles_highfreq_pdcoo2_takatsu_2007} .

We then carry out the same procedure for PdCrO$_{2}$ and substract
the resulting scattering rate from the scattering rate extracted from
PdCoO$_{2}$. This would isolate the last two terms that correspond
to Kondo and EME scattering in $\tau_{tr}^{-1}$. By fitting this
excess scattering rate we find the EME parameters $\lambda_{\rm EME}=0.025$
and $\hbar\tilde{\omega}_{0}=45$meV. Since we use the high temperature
approximation for $\chi''$ we do not fit the low temperature regime
of $\Delta_{\tau}=\tau_{\rm{tr,Cr}}^{-1}-\tau_{\rm{tr,Co}}^{-1}$ but only carry
the fit at temperatures above 80K ($J_{H}\sim72K$). Note that the
constant offset between the PdCoO$_{2}$ and the PdCrO$_{2}$ resistivities
is not fitted. This excess scattering rate comes purely from the Kondo
scattering with the dimensionless coupling $\lambda_{K}=4S\left(S+1\right)J_{K}\nu_{0}$
and a saturation scale set by $J_{H}$, where the value of these couplings
is listed above. 

In the above, one can trace the variuous features of $\rho$ for PdCrO$_{2}$
to each of the terms in Eqn.~\ref{eqn:transport_rate}.The Kondo term increases sharply
and completely saturates at a temperature scale set by $2\pi J_{H}$.
This approach to saturation would be responsible for the sublinear
approach to the $T$--linear regime. This sublinear piece dominates
at $T\sim J_{H}$ over the contributions from the phonon and the magnetoelastic
scattering rates. Eventually, both of these contributions catch up
with the Kondo scattering rate and the resistivity becomes $T$--linear
which can be seen from the high temperature limits of $\mathcal{I}_{4}$
and $\mathcal{I}_{3}$ and the expression for $\tau_{tr}^{-1}$. Taking
the high temperature limit, we can determine the coefficient of the
linear in $T$ scattering rate. Using the asymptotic expressions above
and the values of the fitted dimensionless couplings we get 
\begin{align}
\lim_{T\rightarrow\infty}\frac{d}{dT}\tau_{\rm{tr}}^{-1} & =\left(\frac{k_{B}}{\hbar}\right)2\pi\left(\lambda_{\rm el-ph,o}+\lambda_{\rm el-ph,a}+2\lambda_{\rm EME}\right)\\
 & \sim0.75\left(\frac{k_{B}}{\hbar}\right)
\end{align}
which is close to the Planckian slope $k_B/\hbar$.  

\begin{figure}[h]
\centering
\includegraphics[width=0.45\textwidth]{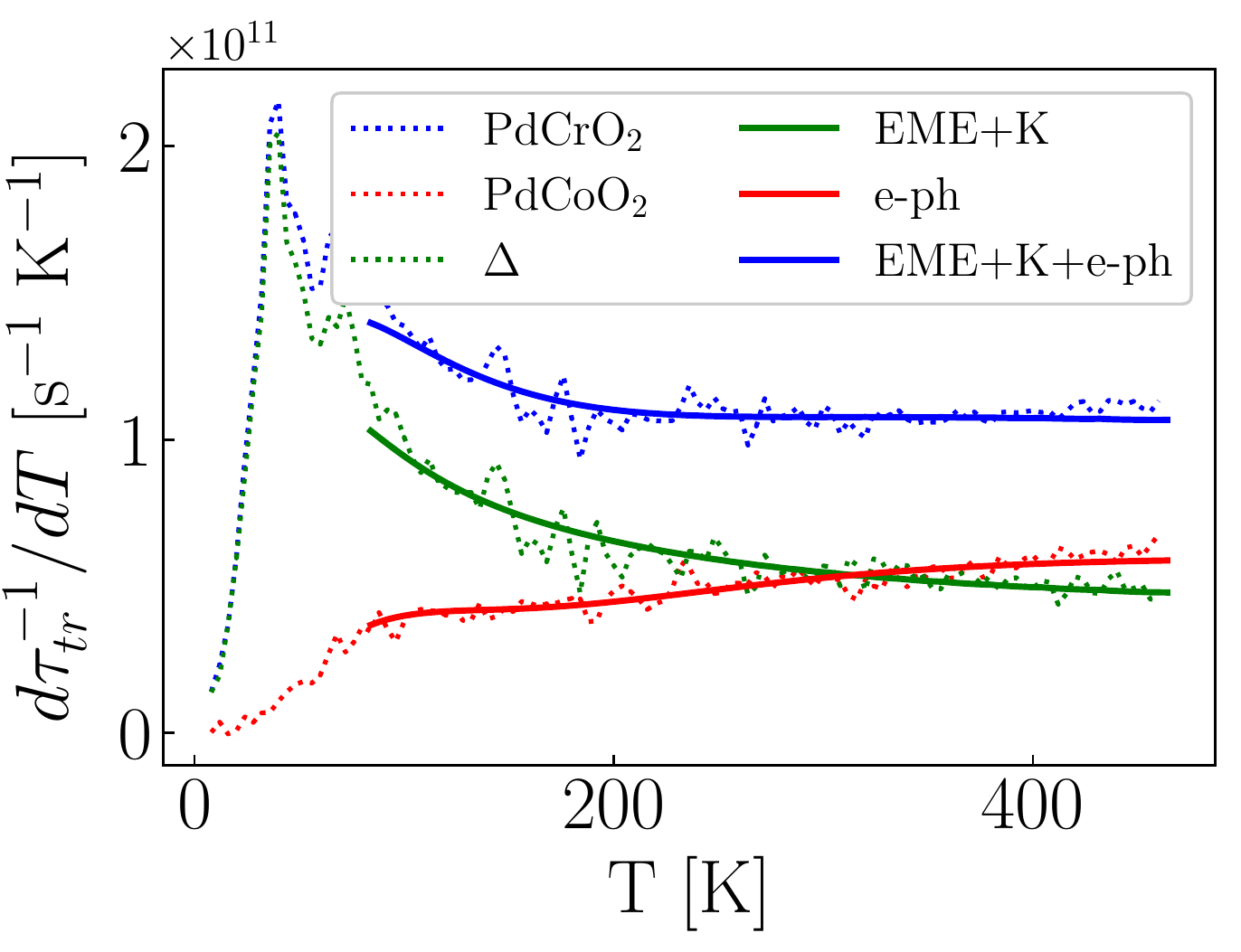} 
\caption{Rate of change of the scattering rates from fit and experiments. In the case where the resistivity is completely saturated we expect this quantity to be constant. However, neither the electron-phonon contribution, nor the Kondo and EME contributions are linear. Nevertheless, the sum of the scattering rates is almost linear, as evidenced by the constant rate of change. }
\label{fig:gammatroverT}
\end{figure}

Finally, we note that neither $\tau_{\rm el-ph}^{-1}$ nor $\tau_{\rm K}^{-1}$ are in the fully saturated regime for the values of the fit parameters
obtained from the above procedure (see Fig.~\ref{fig:gammatroverT}). In contrast, the excess resistivity from the EME
term is generated by a mode that saturates at relatively low temperatures (since the out-of-plane vibration in PdCrO$_2$ turns out to be softer than the in-plane one). At the same time, the addition of the pure Kondo scattering gives
an additional $T$-sublinear contribution to the resistivity. In
total, the $T$-superlinear scattering rate from electron-phonon scattering
and the $T$-sublinear contribution from the Kondo term, together with the $T$-linear EME contribution yield an approximately $T$--linear resistivity for PdCrO$_{2}$.

\section{Out-of-plane resistivity} \label{appendix_caxis}

The starting point of our discussion is a slightly modified version of the model introduced in the discussion of in-plane transport. Here, we assume that in addition to the regular in-plane hopping in the Pd layer, we have a small but non negligible hopping $t_{pp}$ between adjacent Pd layers. The value of this hopping parameter is estimated to be $\sim 0.042 t^p_{nn}$ by taking as a reference the dispersion of PdCoO$_2$ as measured by quantum oscillations \cite{Hicks_mobility_2012}. Once this hopping has been introduced, the current operator in the $z$ direction can be readily calculated. This operator has three distinct contributions (ignoring for simplicity the phonon-assisted, but local-moment independent, term): 
\beq
\bm{J}^{0}&=&-i\sum_{ij\sigma}t^{p}_{ij}\left(\bm{r}_{i}-\bm{r}_{j}\right)p_{i\sigma}^{\dagger}p_{j\sigma}\\
\bm{J}^{1}&=&-i\frac{J_K}{2}\sum_{ilj\beta\alpha}\left(\bm{r}_{i}-\bm{r}_{j}\right)p_{i\alpha}^{\dagger}\left(\bm{S}_{l}\cdot\bm{\sigma}_{\alpha\beta}\right)p_{j\beta}\\
\bm{J}^{2}&=&i\frac{\eta J_K}{2}\sum_{ilj\beta\alpha}\left(\varphi^{(O)}_{il}+\varphi_{jl}^{(O)}\right)\nn \\ 
& &\,\,\,\,\,\,\,\,\,\,\,\,\,\,\,\,\,\,\,\, \times\left(\bm{r}_{i}-\bm{r}_{j}\right) 
 p_{i\alpha}^{\dagger}\left(\bm{S}_{l}\cdot\bm{\sigma}_{\alpha\beta}\right)p_{j\beta}
 \label{eqn:currents_caxis}
\eeq

where $\bm{J}^{0}$ is the usual current operator that has a $z$ component induced by $t_{pp}$, $\bm{J}^{1}$ is an assisted hopping current mediated by a local moment $\bm{S}_l$ at site $\bm{r}_l$ in the Cr layer, and $\bm{J}^{2}$ is a phonon and local moment assisted hopping, with $\varphi^{(O)}_{il}$ being the bond stretching displacement between sites at $\bm{r}_i$ in a Pd layer and $\bm{r}_l$ in a Cr layer. 
For the in-plane transport, these conduction channels are usually shorted by the coherent current $\bm{J}^{0}$ due to the large in-plane Fermi velocity, which is why they were absent from our previous discussions. However, these terms are relevant in the discussion of c-axis resistivity, since the hopping integral $t_{pp}$ is now of similar magnitude as the Kondo and EME couplings which set the overall scale of the incoherent currents.

\begin{figure}[t]
\centering
\includegraphics[width=0.45\textwidth]{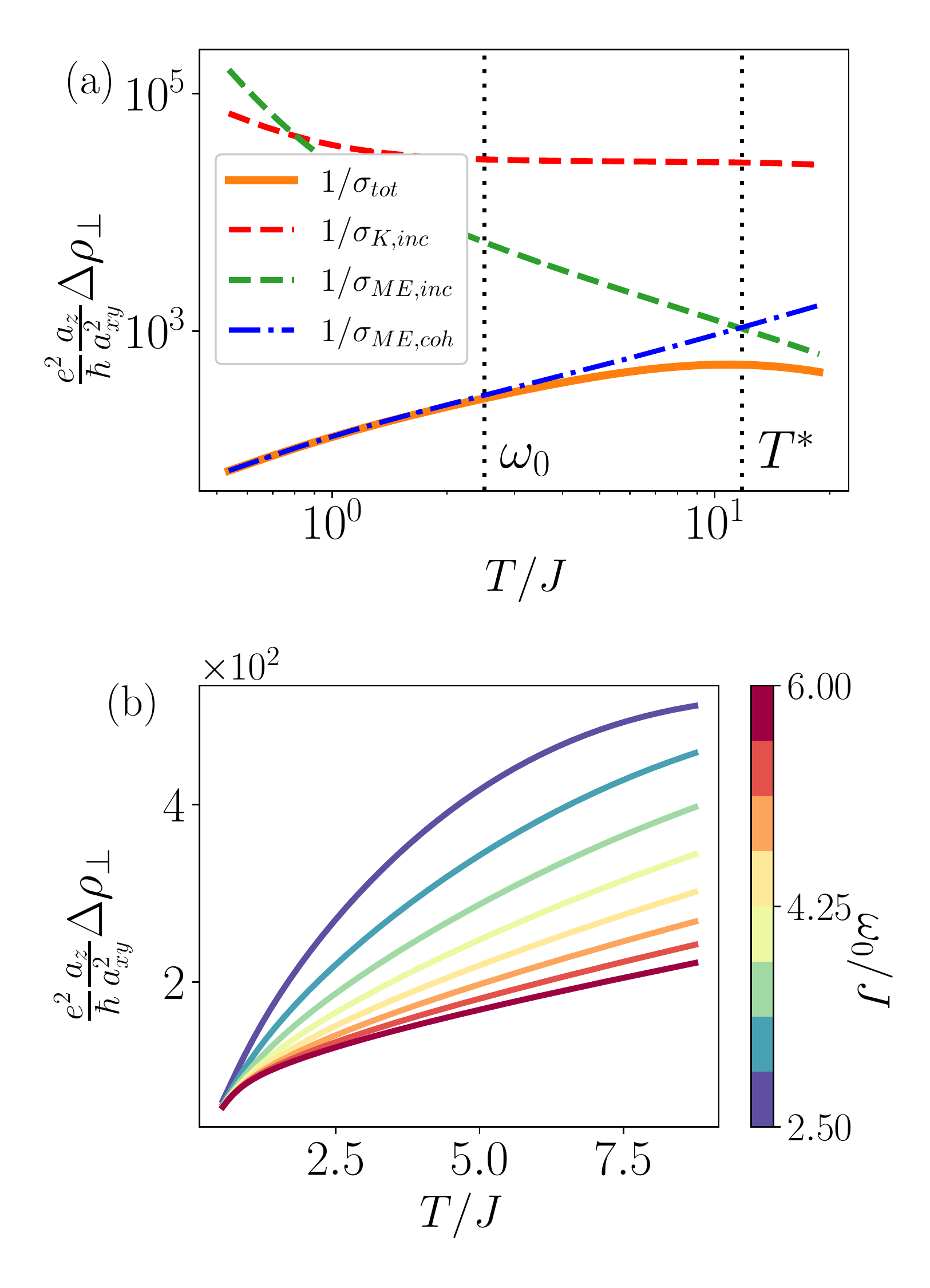} 
\caption{c-axis transport: (a) different contributions to the c-axis resistivity for a soft phonon frequency $\omega_0 =2.5 J_H$. Coherent transport dominates over a broad temperature regime but deviates from $T$ linearity at a temperature $T_*$ set by the electron velocity and the magnetoelastic coupling (see main text). The pure incoherent Kondo scattering is always shorted by other conducton channels. (b) Dependence of the c-axis resistivity on variations of the phonon frequency. As $\omega_0$ becomes softer the incoherent conductivty is enhanced and the coherent scattering increases making a saturation regime accessible at high temperatures. }
\label{fig:outofplane}
\end{figure}

To calculate the conductivity we compute the current-current correlation for the total current $\bm{J}=\sum_{i}\bm{J}^{i}$. Since we work perturbatively in the Kondo and EME couplings, as well as in the out-of-plane hopping $t_{pp}$, the cross correlations between different current operators are subleading and we obtain 
\beq
\sigma^{c} &=& \sigma_{\text{inc}}+\sigma_{\text{coh}}\\
\sigma^{c}_{\text{inc}}&=& \lim_{\omega\rightarrow0}\lim_{\bm{q}\rightarrow0}\sum_{i=1,2}\frac{-1}{\beta V}\langle J^{i}_{z}(\tilde{q})J^{i}_{z}(-\tilde{q}) \rangle\\
\sigma^{c}_{\text{coh}}&=& 4 e^2 t_{pp}^2 a_z^2 \nu_0 \langle \tau_{\rm{tr}} \sin^2 k_z a_z  \rangle
\label{eqn:conducts_caxis}
\eeq

where $\langle \cdot \rangle $ denotes the momentum average over the Fermi surface, and $\nu_0$ is the density of states at the Fermi level. The behavior of each individual contribution to the conductivity is shown in Fig.~\ref{fig:outofplane}(a). Here we see that the incoherent contribution from the Kondo current $\bm{J}^1$ to the resistivity decreases with increasing $T$
and then saturates at a scale set by $J_{\rm H}$. In contrast the contribution of the EME current $\bm{J}^2$ to the resistivity continues to decrease with temperature with $\rho_{\text{inc, EME}}\sim T^{-1}$. At the same time the coherent contribution rises with temperature $\rho_{\text{coh}}\sim T$. We then obtain the total resistivity from parallel addition of each individual conduction channel. Interestingly, we note that there is a crossover temperature scale $T_*$ at which the incoherent and coherent resistivities coincide. At this scale, we expect resistivity saturation. At asymptotically high temperatures, there is an eventual change of trend where the resistivity decreases as a function of temperature as a consequence of incoherent processes. 

To estimate the scale at which this crossover occurs, we simply equate the high temperature limits of $\sigma^c_{\text{inc}}$ and $\sigma^c_{\text{coh}}$. Since the incoherent conductivity is dominated by the EME current correlations, we can approximate at high temperatures:
\beq
\sigma^c_{\text{inc}}\sim \left(\frac{a_{z}}{a_{xy}^{2}}\frac{e^{2}}{\hbar}\right)\lambda_{\rm{EME}}\nu_{0}T
\label{eqn:hight_inch}
\eeq
In the same regime, the coherent contribution to the conductivity can be well approximated by:
\beq
\sigma^c_{\text{coh}} \sim  \left(\frac{a_{z}}{a_{xy}^{2}}\frac{e^{2}}{\hbar}\right)\frac{t_{pp}^{2}\nu_{0}}{(\lambda_{\rm{EME}}+\lambda_{\rm{el-ph}}) k_{B}T}.
\label{eqn:hight_coh}
\eeq
Equating these two conductivities and estimating $v_z\sim t_{pp} a_z /\hbar$ we get the crossover temperature 
\begin{equation}
    T_*\sim\frac{v_z G_z}{\sqrt{\lambda_{\rm{EME}}(\lambda_{\rm{EME}}+\lambda_{\rm{el-ph}})} },
    \label{eqn:tstarscale}
\end{equation}
with $G_z=2\pi \hbar /a_z$. Note that the dependence on the density of states is now hidden in $\lambda_{\rm{EME}}$, which also depends on the frequency of the mode that mediates the EME interaction.

We now analyze the effect of pressure on the $c$-axis resistivity.
 Specifically, applying compressive strain along the $c$ direction introduces changes in the lattice constants, the hopping integrals, the Kondo coupling, and the phonon frequency. The expectation is that the phonon frequencies and lattice constants change algebraically while $J_{\rm{K}}$ and $t_{pp}$ change exponentially, so we focus on the latter in what follows. Additionally, we have $J_{\rm{K}} \propto t_{cp}^2/U,$ with $t_{cp}$ setting the scale of the hybridzation between Cr and Pd layers.  Consider the case where $t_{cp}^2$ has an exponential dependence on $a_z$ similar to that of $t_{pp}$. This could happen, for instance, if the inter-layer hopping is controlled by virtual tunneling via the Cr layers, then $t_{pp}$ is also proportional to $t_{cp}^2/U$. As a consequence, $J_{\rm{K}}$ increases at the same rate as $t_{pp}$ when pressure is applied along the $c-$axis. This entails that upon applying pressure, $\sigma^{c}_{\rm inc}$ grows since the overall scale of the incoherent current increased. At the same time, $\sigma^{c}_{\rm{coh}}$  increases since the increase in the $c-$axis velocity is matched by the increase of the scattering rate with decreasing $a_z$ for the Kondo and EME contributions, but not for the el--ph contribution.

These changes to the resistivity also have an effect on the crossover scale $T_*$. With the scaling of the different hopping parameters, we find that $\lambda_{\rm EME}$ scales with $c$-axis pressure in the same way as $v_z^2$. Following Eqn.~\ref{eqn:tstarscale},  the crossover temperature $T_*$ is expected to decrease with increasing pressure. As such, deviations from $T$-linearity in the resistivity are expected to occur at lower temperatures when pressure is applied.

\section{Incoherent contribution to the $c$-axis resistivity} \label{appendix_caxis_inc}

To evaluate the incoherent contribution to the conductivity we start
by Fourier transforming the current operators that are associated
with the Kondo and EME assisted hoppings between Pd layers.
These current operators are given by 
\begin{equation}
J^{1}_{z}\left(\omega_{n}\right)=\frac{1}{\sqrt{N\beta}}\sum_{\begin{array}{c}
\boldsymbol{k}\boldsymbol{k}'\alpha\beta\\
i\nu_{l},i\nu_{n}
\end{array}}f\left(\boldsymbol{k},\boldsymbol{k}'\right)p_{\boldsymbol{k}\alpha}^{\dagger}\left(i\nu_{l}\right)\left[\bm{S}_{\boldsymbol{k}-\boldsymbol{k}'}\left(i\nu_{l}-i\nu_{n}\right)\cdot\bm{\sigma}_{\alpha\beta}\right]p_{\boldsymbol{k}'\beta}\left(i\nu_{n}+i\omega_{n}\right)
\end{equation}
\begin{align}
J^{2}_{z}\left(\omega_{n}\right) & =\frac{1}{N\beta}\sum_{\begin{array}{c}
\boldsymbol{k}\boldsymbol{k}'\boldsymbol{q}\alpha\beta \ell \\
i\nu_{l},i\mu_{m},i\nu_{n}
\end{array}}\Lambda_{\ell}\left(\boldsymbol{k},\boldsymbol{k}'\right)\varphi^{(O)}_{\ell,\boldsymbol{k}-\boldsymbol{k}'-\boldsymbol{q}}\left(i\nu_{l}-i\mu_{m}-i\nu_{n}\right)p_{\boldsymbol{k}\alpha}^{\dagger}\left(i\nu_{l}\right)\left[\bm{S}_{\boldsymbol{q}}\left(i\mu_{m}\right)\cdot\bm{\sigma}_{\alpha\beta}\right]p_{\boldsymbol{k}'\beta}\left(i\nu_{n}+i\omega_{n}\right)
\end{align}
where the bare vertices are given by 
\begin{align}
f\left(\boldsymbol{k},\boldsymbol{k}'\right) & =J_{K}\left(\frac{2a_{z}}{\hbar}\right) \sin\left(k_{z}a_{z}+k_{z}'a_{z}\right),
\end{align}
and 
\begin{align}
\Lambda_{+}\left(\boldsymbol{k},\boldsymbol{k}'\right) & =\Lambda_{-}\left(\boldsymbol{k},\boldsymbol{k}'\right)=-\eta f\left(\boldsymbol{k},\boldsymbol{k}'\right)
\end{align}
we additionally parametrized the magnetoelastic interaction as $\tilde{\alpha}=-J_{k}\eta$
. Note that we have two phonon operators per local moment since there
are two bond streching modes, one that connects to the upper and one
that connects to the lower Pd layers. With these definitions, we proceed
to calculate the current-current correlation functions 
\begin{equation}
\Pi_{\boldsymbol{k}}^{zz}\left(i\omega_{n}\right)=\frac{-1}{\beta V}\sum_{i}\langle J_{z}^{i}(\tilde{q})J_{z}^{i}(-\tilde{q})\rangle=\Pi_{1}+\Pi_{2}
\end{equation}

There are two diagrams that contibute to leading order, these are
respectively:
\begin{fmffile}{diagram}
\begin{eqnarray}
\parbox{35mm}{\begin{fmfgraph}(100,100)\fmfleft{i} \fmfright{o}
   \fmf{dbl_dots}{i,v1} \fmf{dbl_dots}{v2,o}
   \fmf{dashes}{v1,v2}\fmffreeze
   \fmf{plain,left}{v1,v2,v1}
   \end{fmfgraph}} & = & \Pi_1  \\
   \parbox{35mm}{\begin{fmfgraph}(100,100)\fmfleft{i} \fmfright{o}
   \fmf{zigzag}{i,v1} \fmf{zigzag}{v2,o}
   \fmf{dashes,left=.5,tension=0.3}{v1,v2}\fmffreeze
   \fmf{boson,left=.5,tension=0.3}{v2,v1}\fmffreeze
   \fmf{plain,left}{v1,v2,v1}
   \end{fmfgraph}} & = & \Pi_2 
 \end{eqnarray}
 \end{fmffile}
where the dotted line corresponds to the $\boldsymbol{J}^{1}$ current,
the dashed line denotes the spin susceptibility, the zig-zag line is a $\boldsymbol{J}^{2}$
current, and the wiggly line is a phonon propagator. Note that there are no other phonon vertices since the interaction is diagonal in the phonon mode. If we allow for anharmonicities, further corrections to the $\Pi_2$ correlation function will contribute. 

The first diagram above amounts to the following expression:
\begin{equation}
\Pi_{\boldsymbol{k},1}^{zz}\left(i\omega_{n}\right)=\frac{1}{\beta V}\frac{1}{\beta N}\sum_{\boldsymbol{q}\boldsymbol{p}}\sum_{i\nu_{n}iz_{n}}\left|f\left(\boldsymbol{p}+\boldsymbol{q},\boldsymbol{k}+\boldsymbol{q}\right)\right|^{2}S\left(\boldsymbol{p},iz_{n}\right)\mathcal{G}\left(\boldsymbol{p}+\boldsymbol{q},i\nu_{n}+iz_{n}\right)\mathcal{G}\left(\boldsymbol{k}+\boldsymbol{q},i\omega_{n}+i\nu_{n}\right).
\end{equation}
We expect that in the high temperature limit vertex corrections
can be safely ignored in the perturbative regime when quasi-elastic
collisions dominate and there is no momentum structure to the scattering
potentials (since the spins are completely disordered). In this case, we proceed by performing the bosonic and fermionic Matsubara summations:
\begin{align}
\Pi_{\boldsymbol{k},1}^{zz}\left(i\omega_{n}\right) & =\frac{1}{V}\frac{1}{N}\sum_{\boldsymbol{q}\boldsymbol{p}}\left|f\left(\boldsymbol{p}+\boldsymbol{q},\boldsymbol{k}+\boldsymbol{q}\right)\right|^{2}\int\frac{d\epsilon}{\pi}\int\frac{d\nu}{\pi}\int\frac{dz}{\pi}\chi''\left(\boldsymbol{p},z\right)A\left(\boldsymbol{p}+\boldsymbol{q},\nu\right)A\left(\boldsymbol{k}+\boldsymbol{q},\epsilon\right) \nn \\
 & \ \ \ \ \ \ \ \ \ \ \ \ \ \ \ \ \ \ \ \ \times\left(\frac{n_{F}\left(\epsilon\right)-n_{F}\left(\nu-z\right)}{i\omega_{n}-\epsilon+\nu-z}\right)\left(n_{B}\left(z\right)+n_{F}\left(\nu\right)\right).
\end{align}
We see that overall the current correlation function is a convolution
of electron spectral functions with the spin susceptibility. Once
we perform analytic continuation and take the zero frequency limit,
we get 
\begin{equation}
\sigma_{11}^{c}=\frac{\hbar e^{2}}{VN}\sum_{\boldsymbol{q}\boldsymbol{p}}\left|f\left(p_{z},q_{z}\right)\right|^{2}\int\frac{d\epsilon}{\pi}\left[\int\frac{dz}{\pi}\chi''\left(\epsilon-z\right)A\left(\boldsymbol{p},z\right)\left(n_{B}\left(\epsilon-z\right)+n_{F}\left(-z\right)\right)\right]A\left(\boldsymbol{q},\epsilon\right)\left(-\frac{\partial n_{F}\left(\epsilon\right)}{\partial\epsilon}\right),
\end{equation}
where we already took the $\boldsymbol{k}\rightarrow0$ limit and
we used the local approximation on $\chi''$ to simplify the momentum
sums. Since the electron spectral functions satisfy $A\left(\boldsymbol{p},\omega\right)\approx A\left(\left(p_{x},p_{y}\right),\omega\right)$
we can pull the momentum sums over $q_{x,y}$ and $p_{x,y}$ past
the bare vertex, and to leading order we have 
\begin{equation}
\sigma_{11}^{c}=\frac{2a_{z}}{a_{pl}^{2}}\hbar e^{2}\int\frac{dp_{z}}{2\pi}\frac{dq_{z}}{2\pi}\left|f\left(p_{z},q_{z}\right)\right|^{2}\int\frac{d\epsilon}{\pi}\left[\int\frac{dz}{\pi}\chi''\left(\epsilon-z\right)\rho\left(z\right)\left(n_{B}\left(\epsilon-z\right)+n_{F}\left(-z\right)\right)\right]\rho\left(\epsilon\right)\left(-\frac{\partial n_{F}\left(\epsilon\right)}{\partial\epsilon}\right)
\end{equation}
using 
\[
\sigma_{11}^{c}=\left(\frac{2a_{z}}{a_{pl}^{2}}\frac{e^{2}}{\hbar}\right)J_{K}^{2}\int\frac{d\epsilon}{\pi}\rho\left(\epsilon\right)\left(-\frac{\partial n_{F}\left(\epsilon\right)}{\partial\epsilon}\right)\int\frac{dz}{\pi}\rho\left(\epsilon-z\right)\chi''\left(z\right)\left(n_{B}\left(z\right)+n_{F}\left(z-\epsilon\right)\right).
\]
To evaluate this integral we use the density of states evaluated from
the dispersion in the low energy theory of \cite{sunko_probing_2020}.

Now we focus on the EME current correlation functions
\[
\Pi_{\boldsymbol{q},2}^{zz}\left(i\omega_{n}\right)=-\frac{1}{\beta^{3}VN^{2}}\sum_{\begin{array}{c}
\boldsymbol{k}'\boldsymbol{k}\boldsymbol{p}\\
ik_{n}'ik_{n}ip_{n}\\
\ell
\end{array}}\left|\Lambda_{\ell}\left(\boldsymbol{k},\boldsymbol{k}'+\boldsymbol{q}\right)\right|^{2}D\left(\boldsymbol{k}-\boldsymbol{k}'-\boldsymbol{p},ik_{n}-ip_{n}-ik_{n}'\right)S\left(\boldsymbol{p},ip_{n}\right)\mathcal{G}\left(\boldsymbol{k},ik_{n}\right)\mathcal{G}\left(\boldsymbol{k}'+\boldsymbol{q},ik_{n}'+iq_{n}\right).
\]
Note that $D$ is independent of $\ell$ assuming that the frequencies of the two EME modes per unit cell are roughly equal. This is a consequence of the fact that the mass and the frequency of the two optical modes for the outgoing bonds from a local moment in the Cr layer are the same. Once again we can perform all three Matsubara sums, we then perform analytic continuation and take the appropriate DC limit, which gives the following expression for the c-axis conductivity:
\begin{align}
\sigma_{22}^{c} & =-\frac{e^{2}}{VN}\sum_{\boldsymbol{q}\boldsymbol{k}\ell}\int\frac{d\epsilon}{\pi}\frac{d\mu}{\pi}\frac{dz}{\pi}\left|\Lambda_{\ell}\left(\boldsymbol{q},\boldsymbol{k}\right)\right|^{2}B\left(z\right)\chi''\left(\mu\right)A\left(\boldsymbol{q},\epsilon+z+\mu\right)A\left(\boldsymbol{k},\epsilon\right)\times \nn \\
 & \ \ \ \left[n_{B}\left(-z\right)-n_{B}\left(\mu\right)\right]\left[n_{B}\left(z+\mu\right)+n_{F}\left(\epsilon+z+\mu\right)\right]\left(-\frac{\partial n_{F}\left(\epsilon\right)}{\partial\epsilon}\right).
\end{align}
Note that we have assumed to be a dispersion-less Einstein mode, which simplifies the momentum integrations. Once again, in the quasi 2D limit, we can perform the momentum sums first, which approximately converts each factor of the spectral function into a density of states. Finally, we use the phonon spectral function to evaluate one of the frequency integrals, which gives the result: 
\begin{align}
\sigma_{22}^{c} & =\left(\frac{2a_{z}}{a_{pl}^{2}}\frac{e^{2}}{\hbar}\right)\left(\frac{\hbar\eta^{2}}{M\omega_{0}}\right)\left[2J_{K}\right]^{2}\text{csch}\left(\frac{\beta\omega_{0}}{2}\right)\int\frac{d\epsilon}{\pi}\frac{d\mu}{\pi}\rho\left(\epsilon\right)\cosh\left(\frac{\beta\epsilon}{2}\right)\left(-\frac{\partial n_{F}\left(\epsilon\right)}{\partial\epsilon}\right)\times \nn \\
 & S\left(\mu\right)\left[\rho\left(\epsilon+\omega_{0}+\mu\right)\text{sech}\left(\beta\frac{\epsilon+\mu+\omega_{0}}{2}\right)+\rho\left(\epsilon-\omega_{0}+\mu\right)\text{sech}\left(\beta\frac{\epsilon+\mu-\omega_{0}}{2}\right)\right]\left[\frac{\beta\mu}{2}\text{csch}\left(\frac{\beta\mu}{2}\right)\right].
\end{align}
Importantly, the expression above can be simplified in the limit $\beta \omega \rightarrow 0$. In this case, it is straightforward to check that $\sigma^c_{22}\propto T$. 

\section{In-plane resistivity with acoustic phonons}
\label{appendix_acoustic}

In this section, we consider the contribution to the in-plane resistivity due to the EME term enabled scattering off acoustic phonons. A well-defined (i.e., parametrically large) $T^2$-scaling regime for $1/\tau_{\rm EME}$ exists provided $J_{\rm H}\ll \widetilde{T}_{\rm BG}$. We focus on temperatures, $J_{\rm H}\ll T \ll \widetilde{T}_{\rm BG}$, with $\widetilde{T}_{\rm BG} = 2\widetilde{c}k_F$ being the EME Bloch-Gr\"uneisen (BG) temperature. In this regime, we only consider the effects of quasi-elastic scattering off the spins which have approximately local correlations, such that the spin structure factor can be approximated as a constant in momentum space \textit{and} a delta function in frequency space \cite{mcroberts2022intermediatescale}. In other words, phonons provide a source of inelastic scattering while the uncorrelated spins make this scattering isotropic. This is in contrast to the conventional electron-phonon interaction, where the scattering is mainly small-angle up to the BG temperature. A schematic for the scattering processes is shown in Fig.~\ref{fig:eme_acoustic_phonon}.

\begin{figure}[h]
\centering
\includegraphics[width=0.45\textwidth]{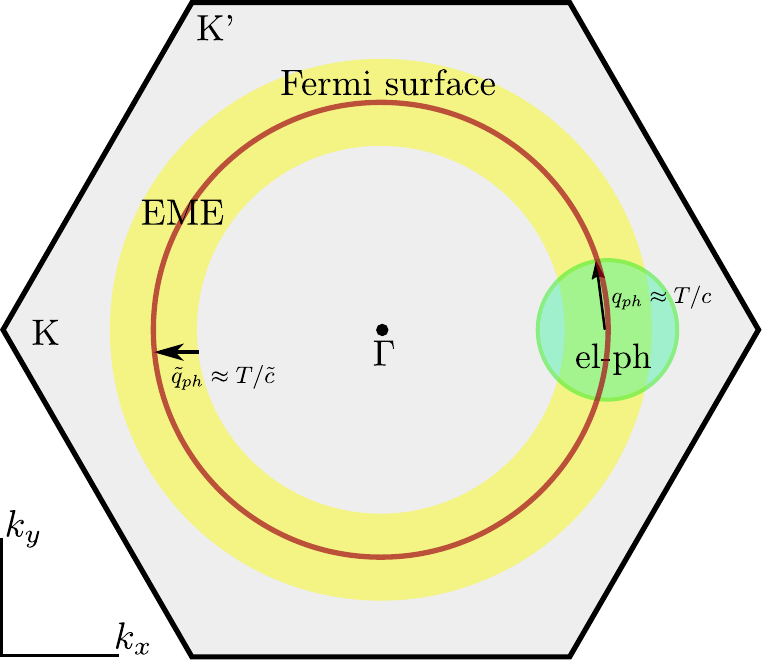} 
\caption{A typical scattering event due to the EME interaction in the local approximation, wherethe spin structure factor is assumed to be featureless, i.e., $S(\omega,\boldsymbol{q}) \approx \rm{const} \times \delta(\omega)$. Starting from $\boldsymbol{k}$ (center of green circle) on the Fermi surface (red circle), the local-moments scatter the momentum uniformly throughout the Brillioun zone via the EME term, such that the annulus (yellow) of phonon modes with momenta $\widetilde{q}_{ph} \leq T/\widetilde{c}$ efficiently scatters electrons isotropically on the Fermi surface. In contrast, the conventional el-ph interaction efficiently scatters electrons to a region enclosed by a sphere of radius $q_{ph} \approx T/c$, as shown in the green circle, such that scattering is mainly small-angle up to $\widetilde{T}_{\rm BG}$. 
}
\label{fig:eme_acoustic_phonon}
\end{figure}

Let us first recall the contribution to the electron-phonon transport scattering time \cite{Allen1993,sadovskii_diagrammatics_2019} within Boltzmann theory (equivalent to Eq.~\eqref{tr_time}), 
\begin{equation}
    \frac{1}{\tau_{\text{el-ph}}}=\frac{4\pi}{\beta}\int_{\Omega}\frac{d\Omega}{\Omega}\alpha^{2}F_{\text{tr}}\left(\Omega\right)\mathcal{K}\left(\beta\Omega\right).
\end{equation}
In the Debye approximation for an acoustic mode in 2D, where $\alpha^{2}F_{\text{tr}}\left(\Omega\right)\approx \lambda_{\text{el-ph}} (3/2)\left(\Omega/\Omega_{D}\right)^{3}\theta\left(\Omega_{D}-\Omega\right)$, there is a factor of $\left(\Omega/\Omega_{D}\right)^{2}$ in the el-ph matrix element due to the weighting of small-angle scattering: $1-\cos\left( \theta_{\boldsymbol{k},\boldsymbol{k'}}\right) \sim q^2 \sim \Omega^2$, where $\boldsymbol{k},\boldsymbol{k'}$ are points on the Fermi surface and $\boldsymbol{q}$ is the phonon wavevector, see e.g., \cite{Ziman_2001}. In contrast, this weighting is absent in the isotropic EME transport time for reasons highlighted above, which can therefore be approximated as 
\begin{equation}
    \frac{1}{\tau_{\text{EME}}}=\frac{4\pi}{\beta}\int_{\Omega}\frac{d\Omega}{\Omega}\alpha^{2}F_{\text{EME}}\left(\Omega\right)\mathcal{K}\left(\beta\Omega\right),
\end{equation}
where $\alpha^{2}F_{\text{\text{EME}}}\left(\Omega\right)\approx\frac{1}{2}\lambda_{\text{EME}}\left(\Omega/\widetilde{\Omega}_{\text{BG}}\right)\theta\left(\widetilde{\Omega}_{\text{BG}}-\Omega\right)$ (note that $\widetilde{\Omega}_{\rm BG} = \widetilde{T}_{\rm BG}$) and $\mathcal{K}$ defined by Eq.~\eqref{def_of_K}. Here we have assumed the BG temperature, rather than the Debye temperature, to be the relevant scale for transport. As described in Fig.~\ref{fig:eme_acoustic_phonon}, the dominant scattering is from thermally excited phonons, which can be seen here by the exponential suppression of $\mathcal{K}(x>1)$. Scattering in this regime if therefore quasi-elastic. In total, we obtain that
\begin{equation}
    \frac{1}{\tau_{\text{EME}}}\approx2\pi\lambda_{\text{EME}}\frac{T^{2}}{\widetilde{T}_{\text{BG}}}, \quad J_{\rm H} \ll T \ll \widetilde{T}_{\rm BG},
\end{equation}
while for $T\gtrsim \widetilde{T}_{\rm BG}$, $1/\tau_{\text{EME}}\approx2\pi\lambda_{\text{EME}}T$, as discussed earlier.

\end{widetext}

\end{document}